\documentclass[trackchanges,twocolumn]{aastex62}
\usepackage{lipsum}
\usepackage{amssymb}
\usepackage{float} \usepackage{listings}
\usepackage{xcolor}
\usepackage{amsmath}
\usepackage{multirow}
\usepackage{booktabs}

\begin{document}
\makeatletter
\let\frontmatter@title@above=\relax
\makeatother

\newcommand\lsim{\mathrel{\rlap{\lower4pt\hbox{\hskip1pt$\sim$}}
\raise1pt\hbox{$<$}}}
\newcommand\gsim{\mathrel{\rlap{\lower4pt\hbox{\hskip1pt$\sim$}}
\raise1pt\hbox{$>$}}}
\newcommand{\CS}[1]{{\color{red} CS: #1}}
\newcommand{\claire}[1]{{\color{orange} Claire: #1}}

\title{\Large 
\textbf{Fast Radio Bursts from White Dwarf Binary Mergers: Isolated and Triple-Induced Channels}}

\shorttitle{FRBs from Binaries and Triples}
\shortauthors{Shariat et al.}

\author[0000-0003-1247-9349]{Cheyanne Shariat}
\affiliation{Department of Astronomy, California Institute of Technology, 1200 East California Boulevard, Pasadena, CA 91125, USA}

\author[0000-0001-9582-881X]{Claire S.\ Ye}
\affiliation{Canadian Institute for Theoretical Astrophysics, University of Toronto, 60 St George Street, Toronto, ON M5S 3H8, Canada}

\author[0000-0002-9802-9279]{Smadar Naoz}
\affiliation{Department of Physics and Astronomy, University of California, Los Angeles, CA 90095, USA}
\affiliation{Mani L. Bhaumik Institute for Theoretical Physics, Department of Physics and Astronomy, UCLA, Los Angeles, CA 90095, USA}

\author[0000-0003-0984-4456]{Sanaea C. Rose}
\affiliation{Center for Interdisciplinary Exploration and Research in Astrophysics (CIERA), Northwestern University, 1800 Sherman Ave,
Evanston, IL 60201, USA}

\correspondingauthor{Cheyanne Shariat}
\email{cshariat@caltech.edu}

\begin{abstract}
The detection of fast radio bursts (FRBs) in both young and old stellar populations suggests multiple formation pathways, beyond just young magnetars from core-collapse supernovae. A promising delayed channel involves the formation of FRB-emitting neutron stars through merger- or accretion-induced collapse of a massive white dwarf (WD). By simulating a realistic stellar population with both binaries and triples, we identify pathways to WD collapse that could produce FRB candidates. We find that (i) triple dynamics open new merger channels inaccessible to isolated binaries, significantly enhancing the overall merger rate; (ii) triple-induced mergers broaden the delay-time distribution, producing long-delay ($\gtrsim1$-8~Gyr) events largely independent of metallicity, alongside a shorter-delay population ($\lesssim100$~Myr) of rapid mergers; (iii) these long delays naturally yield FRBs in older environments such as quiescent host galaxies and galactic halos; (iv) when convolved with the cosmic star-formation history, binary channels track the star-formation rate ($z_{\rm peak} \sim 2$), while triple channels peak later ($z_{\rm peak} \sim 1$), giving a combined local source rate of $R_0 \approx 2\times10^4~{\rm Gpc^{-3}~yr^{-1}}$, consistent with observations; and (v) applying the same framework to Type~Ia supernovae, we find that triples extend the delay-time tail and roughly double the Ia efficiency relative to binaries, yielding rates and redshift evolution in good agreement with observations. If FRBs originate from the collapse of WDs, our results establish triples, alongside binaries, as a crucial and previously overlooked formation pathway whose predicted rates, host demographics, and redshift evolution offer clear tests for upcoming surveys.
\end{abstract}

\section{Introduction}\label{sec:introduction}
Fast radio bursts (FRBs) are millisecond-duration radio transients with dispersion measures (DMs) often exceeding the Galactic contribution, implying extragalactic and even cosmological distances \citep{Lorimer07,Thornton13,Petroff19,CHIME20}. Despite their high event rates \citep[$\sim 10^3$--$10^4$ sky$^{-1}$ day$^{-1}$;][]{Bhandari18,Ravi19,CHIME21}, the nature of FRBs remains unresolved. The detection of a fast radio burst from the Galactic magnetar SGR~1935+2154 \citep{CHIME20,Bochenek20} has established magnetars as a leading channel. However, FRBs exhibit a wide diversity in phenomonology, such as repeating and apparently non-repeating bursts, polarization properties, and host galaxy demographics \citep[e.g.,][]{Bannister19,Prochaska19,Heintz20,Bhandari20,Petroff22,Gordon25,Horowicz25}. These diverse properties strongly suggest that multiple progenitor pathways contribute to FRBs.

FRB progenitor models are often divided into two broad classes: \textit{prompt} and \textit{delayed}. In \textit{prompt} channels, FRBs are powered by compact objects formed shortly after a star formation episode ($\lesssim 100$ Myr), such as young magnetars born in core-collapse supernovae \citep[e.g.,][]{Metzger17,CHIME20,Bochenek20}. In \textit{delayed channels}, the FRB engine forms on longer timescales ($\gtrsim 0.1- 1$ Gyr)
through binary interactions, including accretion-induced collapse (AIC) of massive white dwarfs (WDs), compact object mergers, or other accretion-powered pathways \citep{Totani13,Kashiyama13,Wang16,Deng21,Sridhar21,Rao25}. Host galaxy studies support the presence of both prompt and delayed channels: a majority of FRBs are found in star-forming galaxies
\citep[e.g.,][]{Bassa17,Marcote17,Margalit18,Marcote20,Bhardwaj21_starforming, Niu22,Michilli23,Sharma24, Bruni24, Bruni25,Gordon25,Horowicz25}, while another fraction reside in the outskirts of galaxies or quiescent elliptical galaxies \citep[e.g.,][]{Chatterjee17,Marcote17,Tendulkar17,Fong21,Bhardwaj21_outskirts,Bhandari22,Kirsten22,Ravi22,Michilli23,Sharma23,Shah25,Gordon25,Horowicz25}.  Notably, the association of an FRB source with a globular cluster in M81 \citep{Bhardwaj21_outskirts,Kirsten22} demonstrates that strongly magnetized neutron stars, the likely central engines of FRBs, can form not only from the core collapse of young massive stars, but also from the merger-induced collapse or AIC of old WDs \citep{Kremer21,Lu22,Rao25}.

A promising way to distinguish between these scenarios is through the redshift evolution of the volumetric FRB rate \citep[e.g.,][]{James22,Zhang24,Meng25}. However, current constraints remain weak due to ambiguity between DM and redshift, complex selection biases, and the difficulty in disentangling repeating from non-repeating sources. \citep[e.g.,][]{Hashimoto22,Shin23,Lei25, Paz25}.

Among delayed channels, mergers involving massive WDs are particularly compelling. Both single-degenerate and double-degenerate pathways have been proposed to produce highly magnetized WDs \citep[e.g.,][]{Dessart07,Kashiyama13} or trigger AIC into magnetized neutron stars capable of powering FRBs \citep[e.g.,][]{Nomoto91,Usov92,Shen12,Waxman17,Margalit19,Schwab21,Combi25, Cheong25}. Previous studies have explored binary WD mergers as FRB progenitors \citep[e.g.,][]{Kashiyama13,Lu22,Kremer21,Kremer23,Rao25, Zhang25}, and \citet{Cao18} showed that their predicted rates are among the most consistent with observed FRB statistics. 

However, a large fraction of the stars that evolve into massive WDs ($M_{\rm WD}\gtrsim0.8~M_\odot$), the likely progenitors of magnetized compact objects, originate in hierarchical triple systems. This makes triples a natural and potentially dominant environment for forming FRB progenitors.
On the main sequence, $\sim$35\% of $\sim$2~M$_\odot$ stars and nearly $\sim50\%$ of $\sim$8~M$_\odot$ stars reside in triple star systems \citep{Raghavan2010,Tokovinin14b,Moe17,MoeKratter21,Shariat25_10k,Offner23}. Among triples in the Galactic field, effectively all are observed to be hierarchical \citep[e.g.,][]{Tokovinin14,Tokovinin22_resolvedtriples,Shariat25_10k}, where an inner binary resides on a relatively tighter inner orbit compared to the more distant tertiary star. In this configuration, the outer binary is defined between the tertiary star and the inner binary's center of mass. 
In such hierarchical triples, secular gravitational interactions between the inner and outer orbit give rise to the eccentric Kozai-Lidov (EKL) mechanism \citep{Kozai1962,Lidov1962,Naoz2016}, whereby the inner binary experiences eccentricity-inclination oscillations. EKL oscillations can excite large eccentricities in the inner binary, decreasing its periastron distances, which in turn can lead to tidal circularization, mass transfer, and/or a stellar merger. As a result, triples provide new evolutionary pathways to inner binaries that are inaccessible to isolated binaries. Prior studies have shown that triple dynamics are critical for the formation of accreting WD binaries \citep{Knigge22, Shariat25CV}, ultracompact WD binaries \citep[e.g.,][]{Toonen2016,Toonen20, Shariat23,Shariat25Merge, Rajamuthukumar25}, X-ray binaries \citep[e.g.,][]{NaozLMXB, Shariat24LMXB, Xuan25}, and compact object mergers including neutron stars and black holes \citep[e.g.,][]{Antonini14,Thompson2011,Shappee13,Toonen18,Fragione19a,Fragione19b,Stegmann22,Stegmann25,Perets25}. These findings cement triples as a necessary component for modeling compact object formation and evolution.

In this {\it Letter}, we present the first systematic study of FRB candidates formed from WD mergers in a stellar population including both isolated binaries and hierarchical triples\footnote{see also \citet{Decoene21}, who discuss accreting surrounding debris into a central compact object to power FRB emission through EKL dynamics}. We compare the population demographics of triples -- such as delay time distributions and source rates -- to isolated binary simulations to predict their respective roles in the formation of FRB candidates. The remainder of this study is organized as follows: Section \ref{sec:methodology} outlines the methodology, including the setup of our binary and triple simulations. In Section \ref{sec:results} we highlight the key results, including the triple formation pathways (Section \ref{subsec:formation_channels}), delay time distributions (Section \ref{subsec:delay_times}), and redshift evolution (Section \ref{subsec:frb_redshift}). Lastly, we summarize the main conclusions in Section \ref{sec:conclusions}. A discussion on Type Ia supernovae and supplementary tables is provided in Appendix \ref{app:ia}.

\section{Methodology}\label{sec:methodology}

\subsection{Triple Population and Initial Conditions}
Throughout this paper, we consider hierarchical triple star systems with two stars in the inner binary (masses $m_1$, $m_2$) and a more distant third (tertiary) star (mass $m_3$) on an outer orbit. The inner (outer) binary has a semi-major axis, period, eccentricity, and mass ratio of $a_{\rm in}$ ($a_{\rm out}$), $P_{\rm in}$ ($P_{\rm out}$), $e_{\rm in}$ ($e_{\rm out}$), and $q_{\rm in}=m_2/m_1$ ($q_{\rm out}=m_3/(m_1+m_2)$). The inner binary mass ratio is defined as $q_{\rm in}=m_1/m_2$, where $m_1$ is the {\it initially} more massive star. The mutual inclination of the two orbits is given by $i_{\rm mut}$.

We adopt realistic initial conditions for our stellar triple populations based on the empirical distributions derived by \citet{Shariat25_10k}, which are anchored in \textit{Gaia} observations. In brief, the inner binary properties ($M_1$, $q_{\rm in}$, $P_{\rm in}$, $e_{\rm in}$) are drawn covariantly from the distributions of \citet{Moe17}. The outer orbit follows a log-normal period distribution, thermal eccentricity distribution, and a $q_{\rm out}$ drawn from a power law with slope $-1.4$. Mutual inclinations are assumed isotropic, and all triples are made hierarchical and dynamically stable. Specifically, we require both the octupole hierarchical criterion \citep{Naoz2013sec}
\begin{equation}
\epsilon = \frac{a_1}{a_2} \frac{e_2}{1-e_2^2} < 0.1,
\end{equation}
and the stability criterion \citep{MA2001}:
\begin{equation}
\frac{a_2}{a_1} > 2.8 \left(1+\frac{m_3}{m_1+m_2}\right)^{2/5}\frac{(1+e_2)^{2/5}}{(1-e_2)^{6/5}}
\end{equation}
are satisfied.
All of the above are consistent with the resolved triple population, as observed by {\it Gaia} \citep{Shariat25_10k}. Moreover, the sampling results in multiplicity fractions, i.e., relative fractions of singles, binaries, and triples, that are consistent with current observations \citep{Moe17,Offner23,Shariat25_10k}.

\subsection{Secular Evolution and Stellar Interactions}
For each triple, we solve the hierarchical, secular three-body equations of motion up to the octupole level\footnote{We neglect the hexadecapole \citep[$(a_1/a_2)^5$;][]{Will17, Will21,Conway24} and higher-order terms. Although such corrections can be important in some regimes \citep[e.g.,][]{Holzknecht25,Naoz25}, they are expected to be subdominant here, given the wide outer orbits and typically unequal component masses of our systems.}
of approximation \citep[see][]{Naoz2016}, including the 1st Post Newtonian term for general relativity precession \citep[][]{Ford00,Naoz2013GR}. Tidal evolution is modeled using the equilibrium tides\footnote{Note that here we adopt the equilibrium tide for both stars and WDs. They often tend to underestimate the efficiency of the
tides compared to chaotic dynamical tides for sufficiently large eccentricity \citep[e.g.,][]{Vick19}. Further, considering tight WD+WD binaries, dynamical tides also play an important role in the final stages of circularization and shrinking \citep[e.g.,][]{Fuller12,Vick17,Su22,Xuan25}. Nonetheless, because the main driver of the dynamics takes place during the point-mass stage, we expect that the details of the tides would not qualitatively alter the results.} formalism \citep{Hut80, Eggleton98}, with different prescriptions for convective and radiative envelopes based on the stellar type \citep{Zahn77,Stephan16}.

Single stellar evolution is included for all three stars in the triple using {\tt SSE} \citep{SSE}. The long-term evolution was tested in previous studies \citep[e.g.,][]{NaozLMXB,Stephan16,Angelo22,Shariat23,Shariat25Merge}. Binaries that begin mass transfer are removed from the triple evolution and are then followed 
using the \texttt{COSMIC} binary evolution code \citep{Breivik20}, which also uses {\tt SSE} prescription \citep{BSE}.
Since short-range processes such as tides and mass transfer operate on timescales much shorter than the secular perturbations from the tertiary, the subsequent evolution can be accurately treated using isolated binary evolution. The settings used in \texttt{COSMIC} are set to the defaults, with the following changes:
${\tt qcflag} = 4 $,
${\tt don\_lim} = -2 $.
${\tt qcflag}$ determines the critical mass ratio for unstable mass transfer during Roche Lobe overflow, where setting ${\tt qcflag} = 4 $ follows the prescription of \citet{Belczynski08}, except for WD donors, which follow \citet{BSE}. Setting ${\tt don\_lim} = -2 $ assumes a donor mass loss rate following \citet{Claeys14}. 

Each system is evolved for a random time chosen from $\mathcal{U}(0,12.5)$~Gyr to model a constant star formation history, representative of the local Milky Way population. Note that this choice is mostly inconsequential to our rate estimations since we focus mainly on the delay times: the time it takes to form an FRB candidate after the system forms. We also re-run the same population assuming three different metallicities: $Z_\odot$, $0.1\,Z_\odot$, and $0.01\,Z_\odot$. 

\subsection{Binary Population}
In addition to our triple simulations, we evolve a control population of isolated binaries using {\tt COSMIC}. For this comparison, the inner binaries from our triple sample are re-initialized as stand-alone binaries, evolved with identical stellar evolution prescriptions, and for the same evolution time. This parallel binary population serves two purposes. First, it provides a baseline for assessing how binary evolution alone contributes to potential FRB progenitors. Second, it allows us to isolate the role of secular three-body dynamics in modifying merger pathways. By evolving both populations consistently, we can directly quantify the unique characteristics of triple dynamics relative to isolated binary evolution on the predicted FRB candidate rate and delay time distribution. 

\subsection{FRB Candidate Identification}

The physical origin of FRBs remains uncertain. Their inferred all-sky rates suggest that they are produced by a common class of astrophysical events \citep[e.g.,][]{Petroff19,Ravi22}, yet proposed progenitors span a wide range, from magnetars to various classes of compact object mergers \citep[e.g.,][]{Kashiyama13, Totani13, Wang16, Platts19}. In this work, we adopt an agnostic approach and explore several merger pathways involving WDs that may lead to FRB production. Specifically, our models include three potential FRB formation pathways and one alternative outcome (SNe Ia):

\begin{enumerate}
    \item {\bf FRB Candidate} -- {\it O/Ne+Any:} merger- or accretion-induced collapse of an O/Ne WD with any companion type (degenerate or non-degenerate). 
    \item {\bf FRB Candidate} -- {\it WD+WD:} merger between two carbon-oxygen (CO) WDs of any mass.
    \item {\bf FRB Candidate} -- {\it massive WD+WD:} mergers between a massive white dwarf ($M>0.85~{\rm M_\odot}$), either O/Ne or CO, and another CO white dwarf.
    \item {\bf SNe Ia Candidates} -- {\it massive WD mergers:} We adopt two criteria for SNe Ia progenitors: (i) mergers between a massive CO WD ($>0.85~{\rm M_\odot}$) and WD, including CO and He WDs\footnote{Note that He WDs are excluded from our FRB analysis because it is unclear whether they would cause WD collapse or a detonation on the primary WD's surface \citep[e.g.,][]{Dan11,Shen12,Shen14a,Shen14b,Shen25}.}, and (ii) mergers between two CO WDs of any mass.
    We note that FRB candidates are of interest here; see Appendix \ref{app:ia} for details on SNe Ia rates from our models. 
\end{enumerate}

The first channel (O/Ne+Any) corresponds exclusively to super-Chandrasekhar accretion- or merger-induced collapse of O/Ne white dwarfs. This channel is motivated by models in which the collapse of an O/Ne WD produces a rapidly rotating, highly magnetized neutron star that is a plausible FRB engine \citep[e.g.,][but see also the latter for a different outcome]{Nomoto91,Usov92,Schwab21,Combi25, Schwab16}.

The two WD+WD channels (Cases 2 and 3) allow us to explore double-white dwarf mergers more generally. Although CO+CO mergers are often associated with Type Ia supernovae, a fraction may instead avoid thermonuclear explosion and produce either a massive, strongly magnetized WD or undergo collapse into a neutron star: both discussed as potential FRB progenitors \citep{GarciaBerro12,Kashiyama13,Schwab21}. Observationally, magnetic WDs are systematically heavier and spin more rapidly than their non-magnetic counterparts \citep[e.g.,][]{Liebert03,Tout08,Kawka07,GarciaBerro12,Kepler13,Caiazzo21,Jewett24}, possibly supporting the idea that they originate from mergers.

For these reasons, we do not impose a Chandrasekhar-mass requirement on WD+WD mergers in our FRB channels. Nonetheless, the majority of our systems are near-Chandrasekhar: approximately $70\%$ of Case 2 mergers and $100\%$ of Case 3 mergers have total mass exceeding $1.35~{\rm M_\odot}$.

We emphasize that it is not assumed that these mergers necessarily produce FRBs. Instead, we treat them as candidate pathways to quantify their intrinsic rates and delay-time distributions, and compare the results between triples and isolated binaries. By following the full evolutionary tracks and tagging outcomes such as CO+CO, O/Ne+CO, or O/Ne+non-degenerate mergers, we identify the stellar pathways that could, in principle, give rise to FRBs without presuming the mechanism that ultimately powers the bursts.

We also do not attempt to distinguish between repeating and non-repeating FRBs in this work, given the theoretical uncertainties surrounding the two populations. While some authors have proposed that magnetars in binary systems may preferentially power actively repeating FRBs \citep[e.g., magnetars with aligned spin and magnetic axes in binaries can exhibit sustained activity;][]{Zhang25}, the connection between specific progenitor channels (such as WD mergers or accretion-induced collapse) and repeater vs. non-repeater behavior is not yet established and remains an open question. 

\section{Results}\label{sec:results}

\subsection{Formation Channels of FRBs in Triples}\label{subsec:formation_channels}
\begin{figure*}
    \centering
    \includegraphics[width=0.95\textwidth]{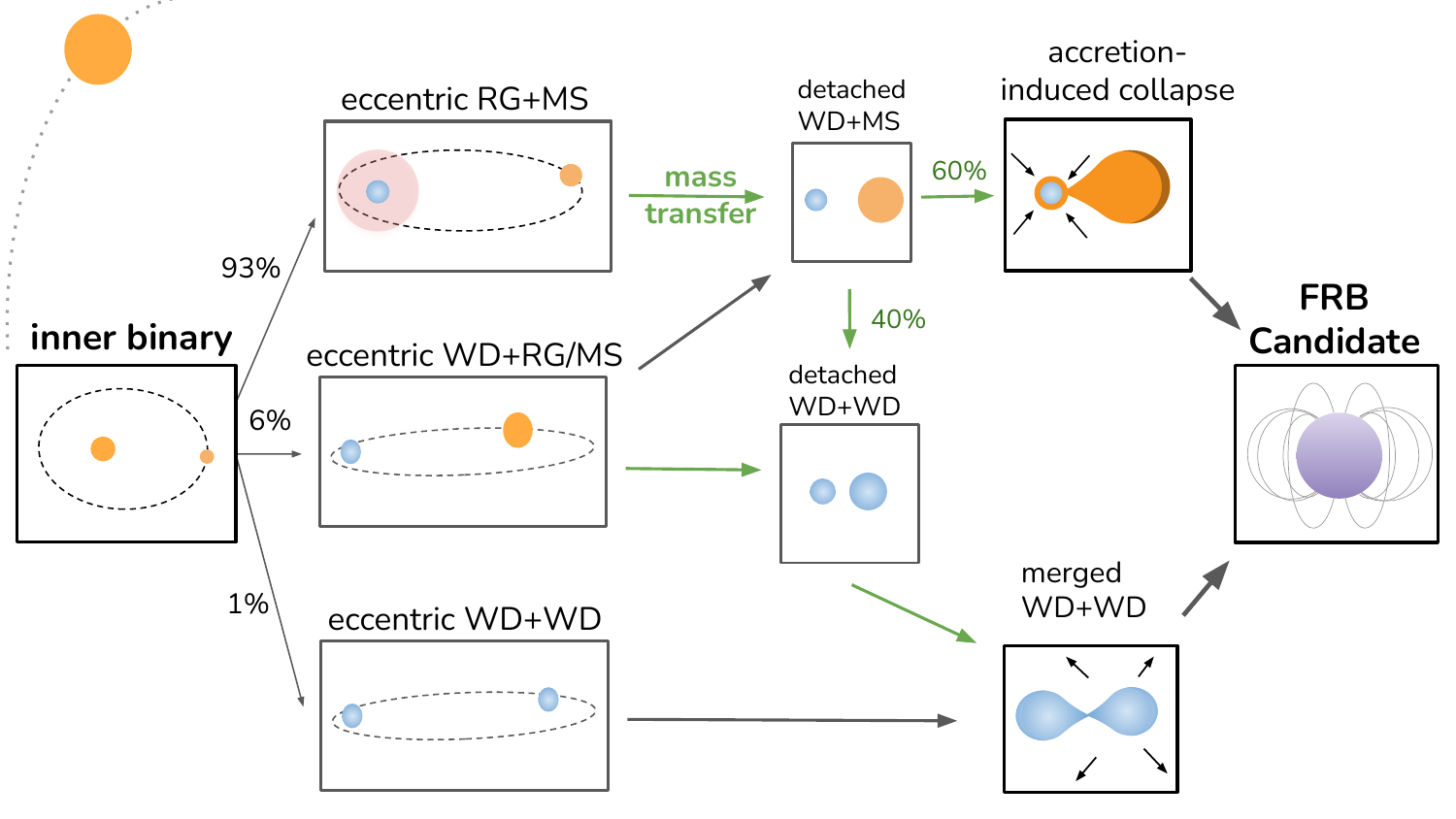}
    \caption{ Possible formation pathways of FRB candidates in triple star systems. Stages of mass transfer (MT), either dynamically stable or unstable, are denoted with green arrows. On the left, we first start with a triple star system containing three main-sequence stars. The first stage of MT in the inner binary can occur when it is an RG+MS, WD+MS/RG, or WD+WD. The evolution follows until the merger of two WDs ($35\%$ of FRB progenitors) or AIC of a white dwarf in a single degenerate scenario ($65\%$). Note that these pathways only represent systems that will become FRB candidates. Broader triple outcomes and their relative fractions are described in Shariat et al. (in prep).
    }\label{fig:FRBs_in_Triples} 
\end{figure*}

After evolving the triple population, we track those that become one of the aforementioned FRB progenitor candidates\footnote{Note that the full set of outcomes from this evolution is discussed in greater detail in Shariat et al. in prep.}. 
Figure~\ref{fig:FRBs_in_Triples} illustrates the formation pathways of FRBs in triple star systems, as observed in our simulated population. An initial inner binary at the zero-age main sequence (ZAMS, $t=0$) is split into three main evolutionary branches based on when mass transfer first begins. The fractions indicate their percentage contribution relative to the entire simulated FRB candidate population. 
\begin{enumerate}
    \item {\it eccentric RG+MS ($93\%$):} The most common scenario is that mass transfer first initiates in the inner binary when the primary evolves to become a red giant (RG)\footnote{We use the class `RG' liberally to encompass various post-main sequence stars including subgiants, red giants, and AGB stars.}. The RG+MS mass transfer can either be stable or unstable, the latter of which leads to a common envelope stage. If the binary does not merge, the result of mass transfer will be a detached WD+MS binary, which can undergo another phase of mass transfer, either stable or unstable. During stable mass transfer, the system would be a cataclysmic variable (CV), at which point it could potentially undergo AIC, making it an FRB candidate. Otherwise, both stable or unstable WD+MS mass transfer could result in a tight, detached WD+WD binary, which could merge and become an FRB candidate. 
    \item {\it eccentric WD+RG/MS ($6\%$):} The second most common outcome is when mass transfer first occurs between a WD+MS or WD+RG inner binary. In this case, the WD evolved independently in an initially wide inner binary ($\gsim10$~au), producing a wide detached WD+MS inner binary. Extreme eccentricities caused by EKL dynamics led to a small periastron separation, allowing tides to shrink and circularize the orbit. Later on, a phase of WD+MS mass transfer begins, resulting in either AIC if the WD retains enough mass to surpass the Chandrasekhar limit or a close WD+WD binary that eventually merges due to gravitational waves, both of which make them FRB candidates.
    \item {\it eccentric WD+WD ($\sim$1\%):} This channel involves head-on collisions of wide WD+WD binaries driven to extreme eccentricities by octupole-level EKL oscillations. In these systems, an initially wide inner binary ($a_{\rm in}\gtrsim10$~au) reaches $e_{\rm in}\gsim0.999$, reducing the periastron to $\lsim10^{-2}$~au and triggering a direct collision. Although eccentric WD+WD mergers represent only a small fraction ($\sim$1\%) of the overall merger population, they contribute significantly by extending the delay times of WD+WD mergers in the overall population (often several Gyr; Section \ref{subsec:delay_times}). This channel is therefore an important contributor to the long-delay merger population in FRBs, and has also been studied in the context of Type~Ia supernovae \citep[e.g.,][]{Toonen18}.
\end{enumerate}

The excitation of eccentricity via EKL, followed by tidal migration 
also delays the onset of Roche-lobe overflow in many RG+MS systems until the red giant is more evolved. In such cases, mass transfer begins at wider orbital separations with greater orbital energy, which can sometimes stabilize otherwise unstable interactions and allow the binary to survive. This mechanism, therefore, not only alters the stability of mass transfer but also extends the delay time for FRB candidate formation relative to isolated binaries.

Compared to isolated binaries, mass-transfer interactions between RG+MS or WD+RG/MS systems occur more frequently in triples. This enhancement arises because wide inner binaries in hierarchical triples unavoidably undergo eccentricity oscillations via the EKL mechanism, which can shrink their periastron distances over time. As a result, binaries that would have otherwise remained detached can experience Roche-lobe overflow, either before or after one component evolves into a WD. In RG+MS systems, this process delays the onset of Roche-lobe overflow until the RG is more evolved or is even on the asymptotic giant branch (AGB).
Starting mass transfer later and at typically wider orbital separations aids in surviving common envelope evolution, which increases the rate of FRB candidates in triples and extends their delay times relative to isolated binaries. 

Considering all FRB progenitor classes together, we find that the merger of two WDs contributes to $35\%$ of FRB progenitors and AIC in a single degenerate scenario  contributes $65\%$. Note that for the rest of this work, the three classes are considered individually.

Table \ref{tab:fractions} summarizes the formation of different FRB progenitor candidates in both binary and triples. The values correspond to the fraction of systems (binaries or triples) formed that produce any given class. For all classes, triples are more efficient at producing WD mergers or AIC. Nearly all predict that $1$ in $1000-10{,}000$ systems form such a merger or collapse event.
\begin{deluxetable}{lcc}
\tablecaption{Fraction of triple and binary stellar systems that evolve into each progenitor channel at solar metallicity. \label{tab:fractions}}
\tablehead{
\colhead{Channel} & 
\colhead{Triple} & 
\colhead{Binary} 
}
\startdata
{\it O/Ne WD+any} & $1.9\times10^{-3}$ & $1.4\times10^{-3}$ \\
{\it WD+WD} & $4.3\times10^{-3}$ & $3.2\times10^{-3}$ \\
{\it massive WD+WD} & $2.3\times10^{-3}$ & $1.6\times10^{-3}$ \\
\enddata
\end{deluxetable}

 Note that some WD mergers could also lead to SNe Ia supernovae. Specifically, mergers involving He WDs (not considered in any of our FRB channels) and/or CO WDs with a massive CO primary ($\gsim0.9~{\rm M_\odot}$) are among the most promising Type~Ia candidates \citep[e.g.,][]{Guillochon10,Ruiter11,Shen18,Shen21,Shen25}. 
 While spatial coincidence between FRBs and Type~Ia supernovae could support the idea of WD mergers as FRB progenitors \citep[e.g.,][]{Kashiyama13}, the most plausible progenitor class considered here involves O/Ne WDs, which are unlikely to produce SNe~Ia.
 Nonetheless, the same triple-induced dynamical processes that enhance FRB candidate formation likewise contribute to the Type~Ia population, which we investigate further in Appendix \ref{app:ia}.

\subsection{Delay Time Distributions}\label{subsec:delay_times}

\begin{figure*}
    \centering
    \includegraphics[width=0.8\textwidth]{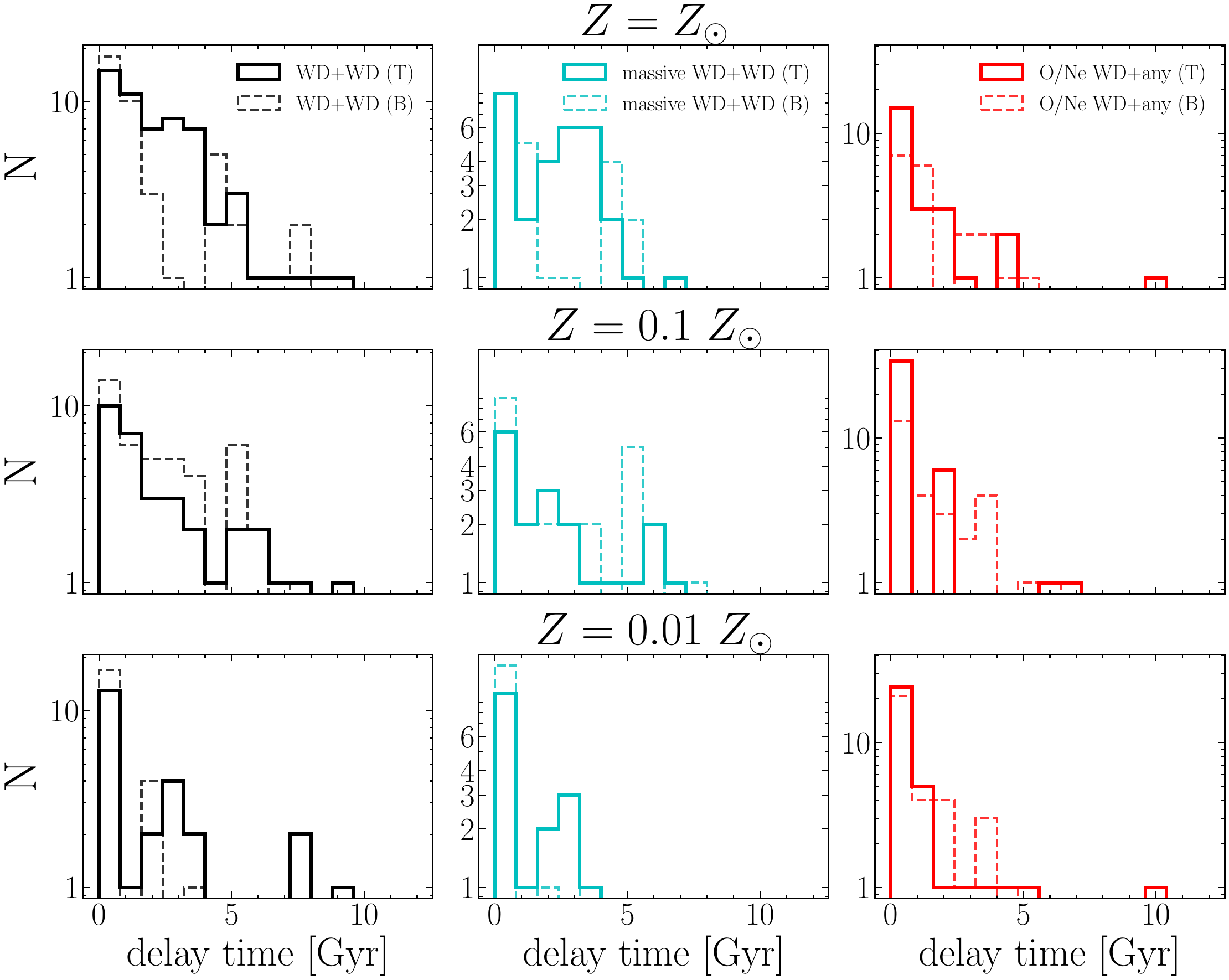}
    \caption{ Predicted delay time distributions of binary mergers in isolated binaries (dashed) and triples (solid) for various metallicities. We show the results for the different FRB progenitor candidates, including all double WD mergers (black), double WD mergers including one massive WD (blue), and O/Ne WD+any (including MS or RG) mergers (red), from left to right. 
    The different rows show different metallicity models.
    }\label{fig:delay_times} 
\end{figure*}

A key discriminant among proposed FRB progenitor models is the delay time distribution (DTD), defined as the time between star formation and the FRB. Models in which FRBs are produced promptly after massive star formation, such as magnetars born in core-collapse supernovae (CCSNe), predict short delay times that closely trace the star formation history (SFH) of their host galaxies \citep[e.g.,][]{Metzger17, Margalit19, Bochenek20, CHIME20}. In contrast, models of delayed channels, such as AIC of massive white dwarfs or compact object mergers, predict longer delay times, leading to a rate that lags behind the SFH \citep[e.g.,][]{Nomoto91, Schwab21, Wang16}. The detection of FRBs in a variety of host galaxies \citep[e.g.,][]{Gordon25,Horowicz25}, including some in quiescent elliptical galaxies \citep[e.g.,][]{Shah25,Eftekhari25} and globular clusters \citep{Bhardwaj21_outskirts,Kirsten22}, lends evidence for a delayed FRB formation channel.

Figure~\ref{fig:delay_times} compares the DTDs for the different FRB progenitor classes and compares those formed in isolated binaries (shaded histograms) to those formed in hierarchical triples (dashed histograms) across three metallicities: $Z=Z_\odot$, $0.1\,Z_\odot$, and $0.01\,Z_\odot$. The three columns correspond to distinct progenitor classes, including double CO WD mergers ({\it left}), mergers containing one massive ($>0.85~{\rm M_\odot}$) WD and another CO WD ({\it middle}), and merger-induced collapse or AIC of an O/Ne WD with any companion ({\it right}).

The left and middle column of Figure~\ref{fig:delay_times} shows that WD+WD mergers from triples exhibit a broader DTD than those from isolated binaries, with a tail extending to $\gtrsim 8\ \mathrm{Gyr}$ at solar metallicity. At solar metallicity, both the total number of WD+WD mergers and their typical delay times are larger in triples than in isolated binaries. 
This arises because many inner binaries begin with wide orbits ($P_{\rm in}\sim10^4$~days) that were unlikely to interact in isolation. In hierarchical triples, however, secular Kozai-Lidov oscillations driven by the tertiary excite the inner binary’s eccentricity. Once the periastron shrinks sufficiently, tidal friction becomes effective, dissipating orbital energy and gradually circularizing and shrinking the orbit to $P_{\rm in}\sim10$~days. This tidal-assisted hardening channel creates new WD+WD binaries that eventually merge, which would not exist in isolated binary evolution. The longer delay times are a natural consequence of this pathway: these systems require multiple stages--first secular eccentricity excitation, then tidal circularization, followed by stellar evolution and mass transfer to the WD+WD stage, then gravitational wave inspiral until merger. On the other hand, some are eccentric WD+WD collisions that occur after a few Gyr, also extending the delay time. In contrast, isolated binaries that merge typically begin much closer, interact earlier through mass transfer, and therefore merge on shorter timescales. Triples thus preferentially populate the long-delay tail of the distribution while also boosting the overall number of WD+WD mergers at solar $Z$.

At very low metallicity (e.g., $\lesssim 0.1 Z_{\odot}$), we find fewer mergers in triples than in isolated binaries across all classes of FRB candidates. This difference arises from both orbital dynamics and stellar evolution. As before, triple inner binaries are often driven to short periods ($P_{\rm in}\lesssim 10$ days) by early EKL cycles followed by tidal circularization. At such close separations, subsequent stellar evolution occurs under conditions very different from solar metallicity. Because reduced line-driven winds at low $Z$ lead to larger pre-WD masses and more compact stellar radii, mass transfer is typically initiated when the mass ratio is extreme ($q \ll 1$). A smaller mass ratio makes unstable mass transfer more likely, thus leading to a larger fraction of mergers.
In contrast, isolated binaries at the same metallicity cover a broader range of initial periods ($P_{\rm orb}\in 10-10^3$~days). Many remain wide enough to avoid early tidal shrinkage, initiating mass transfer later, at wider orbits and mass ratios closer to unity, making survival more likely. As a result, while triples at solar $Z$ enhance the WD+WD merger rate, at low $Z$ their early hardening actually suppresses the production of FRB candidate systems. 
However, the triple dynamical channels ($2$ and $3$ in Section \ref{subsec:formation_channels}) crucially remain pronounced at low $Z$ because they are minimally dependent on metallicity, thus contributing to late-time FRB candidate formation in low metallicity environments.

The rightmost column of Figure \ref{fig:delay_times} shows the O/Ne WD+any channel, dominated by WD+MS and WD+RG mergers, which can yield an AIC if sufficient mass is accreted. 
The DTDs in both triples and binaries are strongly weighted toward short delays ($<1$ Gyr) at all metallicities, with a modest tail to longer times due to some O/Ne WD that only accrete when the secondary evolves to become a giant, which requires a few Gyr on average. 

To summarize, triples and binaries predict qualitatively different delay time distributions for WD+WD mergers, with triples generally producing more mergers with longer delay times at all metallicities. 
At high $Z$, tertiary-induced eccentricity excitation and tidal hardening create WD+WD systems that would not merge in isolation, boosting both the total number of events and the delayed tail of the DTD. At low $Z$, however, reduced stellar winds and more massive progenitors make mass transfer less stable, leading to more mergers during common envelope evolution, thus suppressing WD mergers in triples. The surviving systems are dominated by dynamical eccentric collisions, which are rare but extremely delayed, creating an observable signature of the triple channel. Furthermore, the AIC of O/Ne WDs as FRB candidates shows a relatively delayed DTD as well in both binaries and triples (right panel of Figure \ref{fig:delay_times}) because this channel often requires the secondary star to evolve into a giant to fill its Roche Lobe and begin mass transfer onto the O/Ne WD. These metallicity- and dynamics-dependent contrasts underscore the importance of including triple dynamics when modeling FRB progenitors. In the next section, we translate these distributions into volumetric FRB rates by convolving them with the cosmic star formation history, enabling predictions for the redshift evolution of the FRB population.

\subsection{Redshift Evolution}\label{subsec:frb_redshift}
As the number of observed FRBs grows, especially at cosmological distances, their event rates at various redshifts will become increasingly constrained. Similar constraints were made for SNe Ia supernovae \citep[e.g.,][]{Maoz17}, leading to sharp constraints on their progenitor channels \citep[e.g.,][]{Maoz12_review,Maoz12,Liu23Review}.

At present, however, the observed FRB sample is dominated by low-redshift sources and by intrinsic uncertainties, such as the DM-redshift relation, luminosity function, and fraction of repeating versus non-repeating sources. 
These make it challenging to distinguish models that trace the cosmic star formation history from those that assume delayed DTDs \citep[e.g.,][]{Tang23, Lei25, Gupta25,Meng25}. Namely, while the local ($z\approx0$) FRB volumetric rate and luminosity function are jointly constrained to some extent, the overall redshift evolution of this rate remains poorly understood and model-dependent. 

Current and upcoming wide-field radio surveys, notably CHIME/FRB \citep{CHIME21}, DSA-110 \citep{Law24}, and the future DSA-2000 \citep{DSA-2000}, are poised to dramatically improve FRB detection beyond $z\sim0.5$.
We translate the number of FRB candidates formed per system (binary or triple) into a rate assuming typical stellar multiplicities. Namely, we take $f_{\rm binary} = 34\%$ as the percentage {\it of all stars} formed that reside in isolated binaries\footnote{Note this fraction is among all stars, not star systems.}. We also assume that $f_{\rm triple} = 11.6\%$ of all stars reside in triples \citep[e.g.,][and references therein]{Offner23}, derived using a publicly available tool\footnote{\url{https://github.com/cheyanneshariat/gaia_triples}, see also \citet{Shariat25_10k}} that samples mock stellar populations with singles, binaries, and triples consistent with observed multiplicity statistics. Note that this triple fraction is a lower limit because a notable fraction of stellar systems born in triples become disrupted due to stellar mergers \citep[e.g.,][]{Naoz2014,Toonen2016,Toonen20,Shariat25Merge,Kummer25b}, flyby interactions \citep[e.g.,][]{Michaely2020,Shariat23}, and galactic tides \citep[e.g.,][]{Kaib2014,Stegmann24}. 
This sampling is constructed such that the combined population of stars, across singles, binaries, and triples, follows a Kroupa initial mass function \citep{Kroupa2001}.
Thus, the Galactic rate from the triple channel can be estimated using:
\begin{equation}
    \Gamma \approx10^{-3}~{\rm yr}^{-1}\left(\frac{{\rm SFR}}{2~{\rm yr}^{-1}}\right) \left(\frac{f_{\rm triple}}{0.1}\right) \left(\frac{f_{\rm channel}}{5\times10^{-3}}\right) \ ,
\end{equation}
with a similar equation for binary systems by replacing $f_{\rm triple}$ with $f_{\rm binary}$. Here, $f_{\rm channel}$ is the fraction of FRB candidates formed per stellar system (binary or triple), taken from Table \ref{tab:fractions}. SFR is the star formation rate, taken to be $\approx2$ stars per year, assuming a Kroupa IMF and true SFR of $\sim1~{\rm M_\odot~yr^{-1}}$.
If we assume a number density of $0.02~{\rm Mpc^{-3}}$ for Milky-Way-like galaxies, we estimate a local volumetric rate of 
\begin{equation}
    \mathcal{R}_0\approx10^4~{\rm Gpc^{-3}~yr^{-1}} \left(\frac{\Gamma}{10^{-3}~{\rm yr}^{-1}}\right) \ . 
\end{equation}
Note that the above is a heuristic estimation assuming typical values for SFR and $f_{\rm channel}$ from Table \ref{tab:fractions}. Nonetheless, it demonstrates that the typical rates from the triple channel are comparable to the observed local rates, which is discussed further below. The rates of WD mergers are consistent, to within an order-of-magnitude, of empirical estimates for WD merger rates \citep[e.g.,][]{Brown16,Maoz18,Maoz24}.

We translate the full delay time distributions from our binary/triple simulations (Figure \ref{fig:delay_times}) into FRB source rates as a function of redshift by convolving each DTD with the cosmic star formation rate density (SFRD) from \citet{Madau14}:
\begin{equation}
    \psi(z) = 0.015\,\frac{(1+z)^{2.7}}{1 + \left[(1+z)/2.9\right]^{5.6}}~{\rm M_\odot~yr^{-1}~Mpc^{-3}}.
\end{equation}
Additionally, we combine all of the metallicity models by weighting by the cosmic mean metallicity evolution from \citep{Madau17}:
\begin{equation}
    \log_{10}(\bar{Z}/Z_\odot) = 0.153 - 0.074z^{1.34} \ .
\end{equation}

\begin{figure*}
    \centering
    \includegraphics[width=0.75\textwidth]{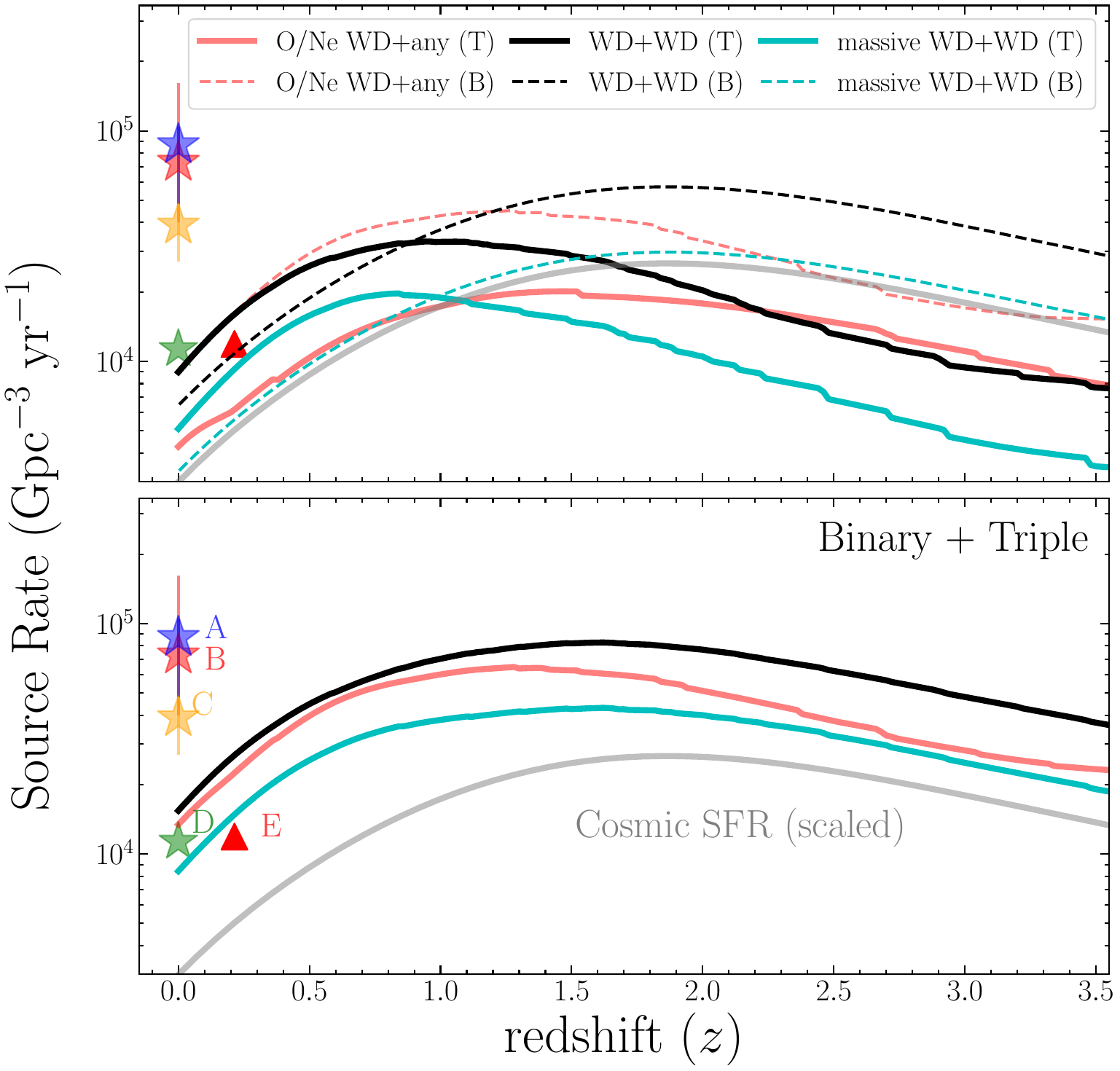}
    \caption{Predicted redshift evolution of FRB source rates in both triples and isolated binaries. The top row shows the distinct contribution from triples (solid lines) and binaries (dashed lines) for different collapse scenarios, while the bottom row combines both channels. In each panel, we show three different FRB progenitor scenarios, including O/Ne WD + any secondary (red), CO WD+WD (black), and massive WD+WD (blue). The bottom panel also contains the cosmic star formation history in gray and various estimates of the local FRB event rate at $z\approx0$, derived by modeling observations. These include \citet{James22} (A), \citet{Shin23} (B), \citet{Meng25} (C), \citet{Chen24} (D), and \citet{Hashimoto22} (E). Note that the rates from our models include only unique FRB candidates, without accounting for repeaters.
    }\label{fig:FRB_rates} 
\end{figure*}

Figure~\ref{fig:FRB_rates} shows the resulting FRB source rates and their redshift evolution for the WD+WD (black), massive WD+WD (cyan), and O/Ne WD+any (red) merger scenarios.
The top panel displays the individual curves for isolated binaries (dashed) and triples (solid), while the bottom panels present the combined rate, assuming a realistic stellar population with both binaries and triples, using the multiplicity fractions described above. 

The binary WD+WD channel broadly tracks the cosmic star formation history (SFH), reflecting their typically short delay times. By contrast, mergers formed in triples are systematically shifted to lower redshift due to longer typical delays ($\gtrsim 1$ Gyr). For instance, the WD+WD merger rate in triples peaks at $z \approx 1.0$, whereas the equivalent binary channel peaks at $z \approx 1.9$. This systematic offset is robust across metallicities (e.g., Figure \ref{fig:delay_times}), with all double-degenerate triple channels peaking near $z \sim 1$. Therefore, if triples contribute significantly to the merger rate, our models predict an observed excess around $z = 1$ relative to binary-only predictions (top panel of Figure \ref{fig:FRB_rates}).

Overall, the different delay signatures are important for interpreting FRB host galaxy demographics: prompt channels (short delays) closely track the cosmic star formation history, and thus should dominate in galactic hosts with recent star formation at $z \lesssim 1$, consistent with some recent host galaxy identifications \citep[e.g.,][]{Bhandari20, Heintz20,Shah25,Eftekhari25,Gordon25}. Moreover, \citet{Gupta25} model the fluence-completeness of the CHIME/FRB Catalog 1 to find that the redshift evolution requires a combination of young and old formation channels.

It is worth emphasizing that the absolute local FRB volumetric event rate remains uncertain, both in the value and its redshift evolution, due to uncertainties in the fluence completeness and DM-redshift relations \citep[e.g.,][]{James22, Meng25}. Nevertheless, combining the DTD analysis with the convolved redshift-dependent rates demonstrates two robust trends: (i) triple evolution broadens the temporal window over which FRBs can occur, shifting a significant fraction to later cosmic times, and (ii) metallicity introduces distinct differences between binary and triple channels. Together, these results provide a natural framework for understanding why FRBs are found in both young, star-forming galaxies and older, passive hosts, and suggest that surveys targeting low-metallicity environments or $z\sim1$ systems will be especially valuable for constraining progenitor models.
Table~\ref{tab:frb_rate_summary} summarizes the main characteristics for FRB rate evolution, including the local rate ($\mathcal{R}_{\rm 0}$), the peak redshift ($z_{\rm peak}$), and the peak rate ($\mathcal{R}_{\rm peak}$).

To compare our predictions with observations, we compile several recent estimates of the FRB volumetric rate (colored points in Figure \ref{fig:FRB_rates}). It is important to note that these rates are not direct measurements. Inferring an event rate requires assumptions about the FRB luminosity function, source population, and survey selection effects, all of which remain highly uncertain. Additional uncertainties arise from the poorly constrained mapping between DM and redshift, and from the treatment of repeating sources, which are either excluded, down-weighted, or counted as single events depending on the analysis. Thus, the published estimates should be viewed as model-dependent and approximate, rather than definitive measurements of the FRB rate. We briefly describe each below.

Different studies using different samples and methodologies yield local FRB rates that vary by up to an order of magnitude. \citet{Chen24}, by fitting 474 apparently non-repeating CHIME/FRB events, find a monotonically declining formation rate $\rho(z) \propto (1+z)^{-4.9\pm0.3}$ with a local rate of $1.1\times10^4~{\rm Gpc^{-3}~yr^{-1}}$. \citet{Meng25}, by incorporating instrumental selection effects with Monte Carlo simulations of CHIME data, report higher rates of $2.3^{+2.4}_{-1.2}\times10^5$ and $3.9^{+2.5}_{-1.2}\times10^4~{\rm Gpc^{-3}~yr^{-1}}$ for pivot energies above $10^{38}$ and $10^{39}$ erg, respectively; we adopt the latter for consistency with other works. \citet{James22} use ASKAP and Parkes to estimate $8.7^{+1.7}_{-3.9}\times10^4~{\rm Gpc^{-3}~yr^{-1}}$ above $10^{39}$ erg. \citet{Shin23} fit fluence and DM distributions from CHIME, finding $7.3^{+8.8}_{-3.8}\times10^4~{\rm Gpc^{-3}~yr^{-1}}$. Finally, \citet{Hashimoto22}, analyzing $164$ non-repeating CHIME events, derive a lower local rate of $1.5\times10^4~{\rm Gpc^{-3}~yr^{-1}}$ at $0.05<z\leq0.3$. 

The above estimations are included in Figure \ref{fig:FRB_rates}. In general, the predicted rates for each scenario reside around $\approx1\times10^4~{\rm Gpc^{-3}~yr^{-1}}$ at $z=0$ and $\approx5\times10^4~{\rm Gpc^{-3}~yr^{-1}}$ at $z\approx1-2$, when summing the contribution from both binaries and triples. These estimates are broadly consistent with observations, particularly \citet{Shin23} and \citet{Hashimoto22}. However, since we report formation rates of candidate FRB progenitors from our simulations, rather than actual {\it event} rates, the unknown fraction of repeating sources complicates the comparison and makes our reported rates strictly lower limits.

Estimates of the magnetar birth rate suggest that magnetars form at $\sim10\%$ of the CCSNe rate \citep[e.g.,][]{Kouveliotou98,Gill07}. Using the CCSNe rate at $0.1<z<0.5$ of $\sim2.5\times10^5~{\rm Gpc^{-3}~yr^{-1}}$ \citep{Dahlen04}, this implies a magnetar formation rate from CCSNe of $R_{\rm mag,CC}\sim2.5\times10^4~{\rm Gpc^{-3}~ yr^{-1}}$. 
If a fraction of these magnetars produce FRBs, their source rate would be comparable to that inferred observationally and to the candidate rates predicted here. Similarly, \citet{Nicholl17}  consider various possible millisecond magnetar production channels to estimate a number density, $\sim 10^4~{\rm Gpc^{-3}~yr^{-1}}$ for FRB emitters. 
An important prediction of the magnetar-from-CCSNe scenario is that FRBs should closely trace recent star formation and star-forming regions within their host galaxies. In contrast, WD-collapse channels predict FRBs in both star-forming galaxies and old stellar environments, reflected from their intrinsically broad delay-time distributions that include both rapid and Gyr-scale components.

In practice, it is also possible that only some fraction of mergers or AIC in an FRB progenitor class is capable of powering an observable FRB. In this case, the quoted rates can be rescaled by an efficiency factor $f_{\rm FRB} \leq 1$ describing the probability that a given merger produces an FRB-emitting remnant. In other words, the current results can be interpreted as an upper limit.

\section{Conclusions}\label{sec:conclusions}

In this work, we explore the formation of FRB progenitors through white dwarf mergers/collapse in binary and triple star systems. Using detailed three-body simulations combined with binary population synthesis, we trace the evolutionary pathways leading to classes of mergers and accretion-induced collapse events, and investigate their implications for the FRB population. Our main results are summarized as follows:
\begin{enumerate}
    \item {\it Formation pathways of FRBs in triples:}  
            Hierarchical triples open new formation pathways to FRB candidates that are inaccessible in isolated binaries (Figure \ref{fig:FRBs_in_Triples}). In particular, eccentricity excitation through the EKL mechanism can shrink wide inner binaries that would otherwise remain detached, producing elevated rates of binary mass  transfer and thus the collapse of super-Chandrasekhar WDs to potential FRB-emitting young neutron stars. At solar metallicity, this channel substantially boosts the total number of mergers relative to isolated binary evolution, while at low metallicity ($Z\lesssim0.1\,Z_{\odot}$), the enhancement is suppressed due to increased unstable mass transfer rates from larger pre-WD masses. However, the dynamically-induced merger channel, where a tertiary triggers a WD+WD or WD+MS collision in the inner binary, is independent of metallicity and provides an important contribution with larger typical delay times ($\gsim1$~Gyr).
    \item {\it Delay times:}  
            The DTDs of the FRB candidates in triples are systematically broader than those from isolated binaries (Figure \ref{fig:delay_times}). At solar metallicity, tertiary-driven tidal shrinkage produces WD+WD mergers on long timescales, extending the DTD tail to $\gtrsim 8$ Gyr. 
            At low metallicity, while the total number of triple mergers decreases, the surviving systems are dominated by eccentric WD+WD collisions, which are independent of metallicity and occur on very long timescales. Indeed, FRB hosts show a weak-to-null metallicity preference  \citep[e.g.,][]{Yamasaki25}.
    \item {\it Host Environments:}  
            Because most FRB candidates formed through secular three-body dynamics have longer average delay times, they are not expected to trace ongoing star formation in all cases. Instead, a substantial fraction will reside among older stellar populations, including quiescent host galaxies or galactic halos. Thus, triple formation can naturally account for the observations of FRBs in a variety of host environments \citep[e.g.,][]{Eftekhari25,Shah25,Gordon25}.
    \item {\it Rates and Redshift evolution:}  
            By convolving the DTDs with the cosmic star formation history, we predict the redshift evolution of the FRB rate (Figure \ref{fig:FRB_rates}). Binary-driven WD+WD merger channels largely track the star formation rate while triple-driven channels are shifted towards lower redshift.
            This systematic offset is a robust prediction of the triple scenario: if WD+WD mergers in triples contribute significantly to FRBs relative to other channels, their observed rate distribution should favor lower redshifts. For the AIC scenario, we find that both binary and triple scenarios lead to moderately delayed distribution. Combining both binary and triple channels across metallicities, our models predict a local FRB candidate source rate $R_0\approx2\times10^4~{\rm Gpc^{-3}~yr^{-1}}$ (Figure \ref{fig:FRB_rates}), comparable to recent observations (see also Table \ref{tab:frb_rate_summary}).
    \item {\it Implications for SNe Ia:}   
            Using the same simulations, we revisited the role of WD mergers in producing SNe~Ia through both binaries and triples. By only considering double-degenerate mergers with a massive CO WD primary (including He WD secondaries), as motivated by detonation models and observations, we find that triples extend the delay-time distribution and yield SNe~Ia at roughly twice the efficiency of isolated binaries ($12\times10^{-4}$ vs.\ $6\times10^{-4}~{\rm SNe~M_\odot^{-1}}$; Figure \ref{fig:Ia_rates_redshift}). Combining both binary and triple channels, we predict an overall efficiency of $\approx2\times10^{-3}~{\rm SNe~M_\odot^{-1}}$, a local volumetric rate of $\approx3\times10^4~{\rm Gpc^{-3}~yr^{-1}}$, and a redshift evolution all in good agreement with observations (Appendix \ref{app:ia}). These results indicate that triples are an essential component of the SNe~Ia progenitor population, although future work is needed to assess the theoretical uncertainties in modeling Type Ia progenitors. 
\end{enumerate}

Our results highlight the distinct role of triple dynamics in forming candidate FRB progenitors.
Testing these predictions observationally offers a promising path to identifying the dominant FRB formation channels. Arcsecond-scale localizations are already constraining host-galaxy properties, while sub-arcsecond positions will continue to pinpoint the local environment of FRBs. Together providing powerful clues to their origins.
The discovery of a nearby stellar companion to an FRB source would further support merger- or accretion-induced collapse pathways \citep[see also][]{Zhang25}. Upcoming wide-field surveys such as CHIME/FRB Catalog~2 \citep{CHIMEFRB_2}, CHORD \citep{CHORD}, and DSA-2000 \citep{DSA-2000} will expand FRB samples, enabling precise measurements of rates, luminosity functions, repeater fractions, and redshift evolution. These advances will provide the statistical power needed to test predictions and to distinguish the relative contributions of binaries, triples, and other FRB progenitor pathways.

\section{Acknowledgments} \label{acknowledgments}
We thank the anonymous referee for constructive feedback as well as Chang Liu, Kaitlyn Shin, Kenzie Nimmo, Barack Zackay, and Dan Maoz for valuable discussions.
We thank the 56th Annual Meeting of the Division on Dynamical Astronomy (DDA) for sparking early conversations about the topic. 

C.S. acknowledges support from the Joshua and Beth Friedman Foundation Fund and the Department of Energy Computational Science Graduate Fellowship. 
This material is based upon work supported by the U.S. Department of Energy, Office of Science, Office of Advanced Scientific Computing Research, under Award Number DE-SC0026073. C.S.Y. acknowledges support from the Natural Sciences and Engineering Research Council of Canada (NSERC) DIS-2022-568580 and the Alfred P. Sloan Foundation. S.N. acknowledges the partial support of NSF-BSF
grant AST-2206428 and NASA XRP grant 80NSSC23K0262,
as well as Howard and Astrid Preston for their generous support.
S.C.R. thanks the CIERA Lindheimer Fellowship for support.

\appendix 
\section{Type Ia Supernovae}\label{app:ia}
Throughout this study, we have considered mergers of two white dwarfs. A natural extension of our results is to the double-degenerate scenario for Type Ia supernovae (SNe Ia). The role of triples in producing SNe Ia has been explored in several works \citep[e.g.,][]{Thompson2011,Katz12,Hamers13,Hamers19,Toonen18,Hallakoun19,Michaely21,Rajamuthukumar23}, and our simulations with realistic initial conditions allow us to revisit this question. Specifically, we focus on the delay times, rates, and redshift evolution of SNe Ia in both binary and triple channels.

We adopt two criteria for SNe Ia progenitors: (i) mergers between a massive CO WD ($>0.85~{\rm M_\odot}$) and another WD, including He WDs (which were excluded from our FRB analysis), and (ii) mergers between two CO WDs of any mass. The first channel is deemed the `Physical' channel as it is motivated by explosion models that find stable detonations in C/O+He WD systems with massive primaries \citep[e.g.,][]{Guillochon10,Ruiter11,Shen18,Shen21}, while the second follows the canonical double-degenerate model.

A recent triple study by \citet{Rajamuthukumar23} found that the SN Ia rate from triples is comparable to that from binaries, both $\sim3\times10^{-4}~{\rm SNe~M_\odot^{-1}}$, which combined is still a factor of a few less than an observed estimate of the local rate $\sim13\times10^{-4}~{\rm SNe~M_\odot^{-1}}$. However, they allow {\it all} CO+He WD mergers to produce SNe Ia, irrespective of mass. In fact, $96-99\%$ of the SNe Ia in their binary/triple evolution models arise from such mergers, making it imperative to their results. This likely overestimates the contribution, since hydrodynamical simulations suggest that only sufficiently massive ($\gsim0.85~{\rm M_\odot}$) primaries can reproduce the observed luminosities of most SNe Ia \citep[e.g.,][]{Pakmor12,Hoeflich17,Shen18}, unlike the typical WD mass of $\sim0.6~{\rm M_\odot}$ while likely dominates the Type Ia progenitors in their simulations. Furthermore, their simulations were truncated at a maximum wall time of 5 hr, which can make it difficult to capture the long-term dynamics that often drive late-time triple-induced mergers.

Guided by these results, we adopt two models. An `Inclusive' model that considers all CO+CO WD mergers as potential SNe Ia progenitors, independent of component masses. The second is a `Physical' model which requires that the primary be a massive CO WD ($>0.85~{\rm M_\odot}$), consistent with explosion models that reproduce the typical luminosities of normal SNe Ia \citep[e.g.,][and references therein]{Shen25}.

\begin{figure*}
    \centering
    \includegraphics[width=0.7\textwidth]{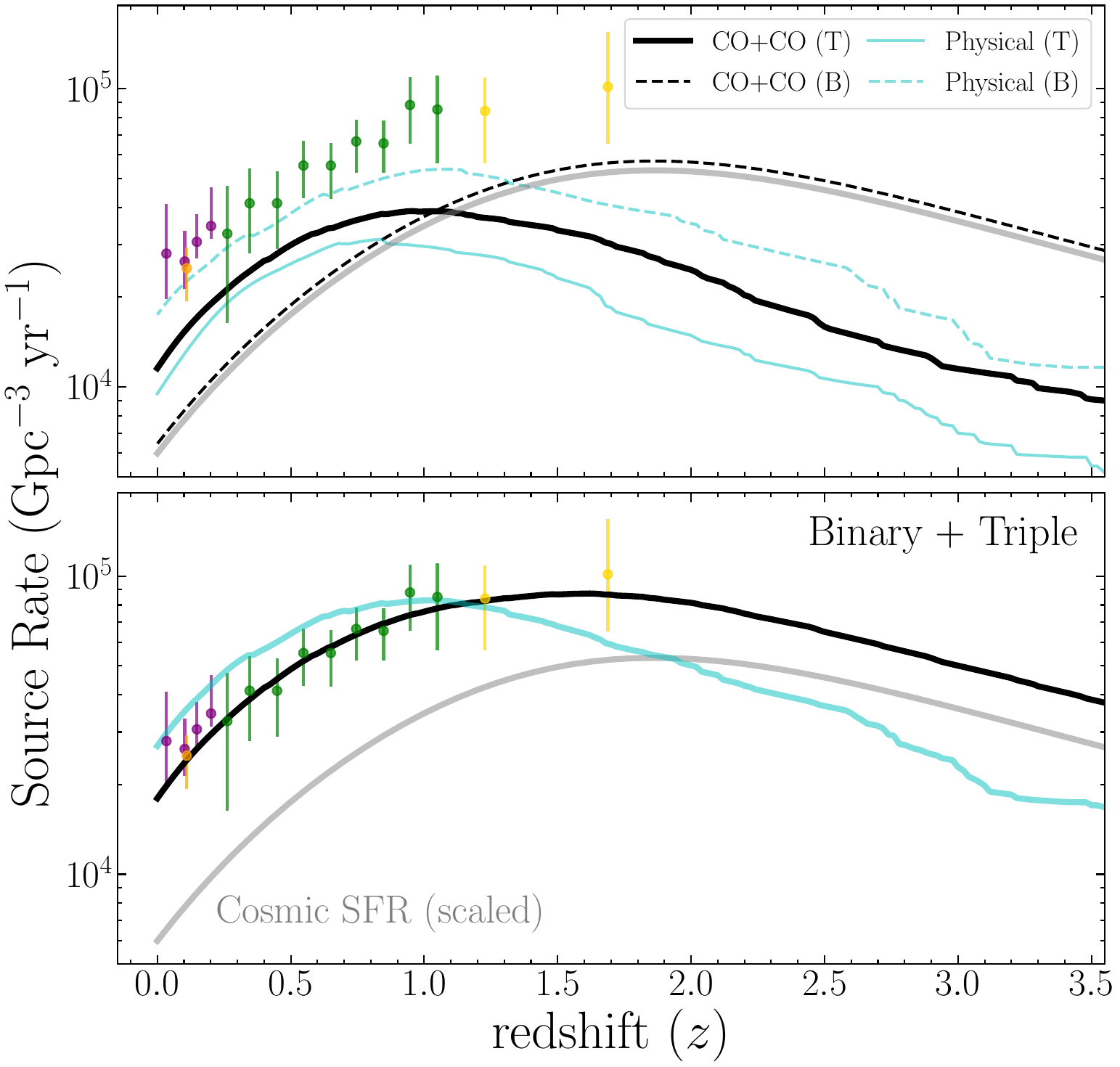}
    \caption{Predicted rates and redshift evolution of Type Ia supernova from a stellar population containing both binaries and triples.
    The top panel shows the individual rates of triples (solid lines) and isolated binaries (dashed lines), while the bottom panel shows their combined rates.
    We assume two different Type Ia progenitor models: one including all CO + CO WD mergers (black), and one that only includes massive CO WD mergers ($>0.9~{\rm M_\odot}$) with either a CO WD or a He WD (blue).
    The bottom panels also displays observed rates from \citet{Dilday10} (purple), \citet{Graur13} (orange), \citet{Perrett12} (green), and \citet{Graur11} (yellow). 
    }\label{fig:Ia_rates_redshift} 
\end{figure*}

In Figure \ref{fig:Ia_rates_redshift}, we show the predicted redshift evolution of the SN Ia rate from binaries and triples. The top panel highlights that in the Inclusive model, binaries typically produce shorter delay times, while triple-induced mergers extend the distribution to longer delays. Namely, despite there being fewer triples than binaries by a factor of 3, triples still contribute more significantly to CO+CO white dwarf mergers at $z\lsim1$ than isolated binaries.

In the Physical model, where only massive CO WDs are included, the shape of the rate evolution is similar in both binaries and triples. The bottom panel, combining binaries and triples in a 3:1 ratio, shows that both models are consistent with observed SN Ia rates across redshift \citep[compiled by][]{Maoz17}. This predicts a local volumetric Ia rate of $\mathcal{R}_{\rm 0,Ia}\approx3\times10^4~{\rm Gpc^{-3}~yr^{-1}}$. 
Importantly, both the estimated rate and redshift dependence of our predicted rates align with current observational constraints. 

In terms of efficiency of Ia formation, we find that the binary channel contributes $6\times10^{-4}~{\rm SNe~M_\odot^{-1}}$ and triples contribute $1.5\times10^{-3}~{\rm SNe~M_\odot^{-1}}$. In total, this leads to an estimate of $\sim2\times10^{-3}~{\rm SNe~M_\odot^{-1}}$, consistent with the observed rate of $(1.3\pm0.2) \times10^{-3}~{\rm SNe~M_\odot^{-1}}$ \citep{Maoz12}. 

Note that modeling SNe Ia progenitors through binary evolution carries substantial theoretical uncertainties, spanning common-envelope physics, mass transfer stability, and explosion mechanisms \citep[e.g.,][]{Claeys14,Liu23Review}. Our results are therefore uncertain. Instead, we demonstrate that, under reasonable assumptions and physically motivated definitions of SNe Ia progenitors, including both binary and triple formation pathways, yields rates and redshift evolution of SNe Ia are broadly consistent with the literature. A more detailed exploration of the binary physics and explosion criteria will be necessary in future work. Moreover, empirical  constraints on the short-period double WD population in our Galaxy will help solidify their contribution to SNe Ia \citep[e.g.,][]{Brown16,Hallakoun19,Maoz18,Maoz24}.

\section{Redshift evolution of FRB candidates}
In Table \ref{tab:frb_rate_summary}, we summarize the properties of the redshift evolution for each FRB candidate class. This includes the local source rate ($\mathcal{R}_0$), the source rate at peak redshift ($\mathcal{R}_{\rm peak}$), and the peak redshift for different metallicities ($z_{\rm peak}$), channels, and stellar systems.

\begin{table*}[h]
\caption{Peak redshift and FRB rates for all channels, modes, and metallicities. Rates are in Gpc$^{-3}$ yr$^{-1}$; `Total' sums triple and binary curves before measuring peaks.}

\label{tab:frb_rate_summary}
\centering
\begin{tabular}{llccccccccc}
\toprule
\multicolumn{2}{c}{} &
\multicolumn{3}{c}{\textbf{Triple}} &
\multicolumn{3}{c}{\textbf{Binary}} &
\multicolumn{3}{c}{\textbf{Total}} \\
\cmidrule(lr){3-5}\cmidrule(lr){6-8}\cmidrule(lr){9-11}
Channel & Metallicity &
$\mathcal{R}_0$ & $\mathcal{R}_{\rm peak}$ & $z_{\rm peak}$ &
$\mathcal{R}_0$ & $\mathcal{R}_{\rm peak}$ & $z_{\rm peak}$ &
$\mathcal{R}_0$ & $\mathcal{R}_{\rm peak}$ & $z_{\rm peak}$ \\
& & \multicolumn{2}{c}{[Gpc$^{-3}$ yr$^{-1}$]} & &
      \multicolumn{2}{c}{[Gpc$^{-3}$ yr$^{-1}$]} & &
      \multicolumn{2}{c}{[Gpc$^{-3}$ yr$^{-1}$]} & \\
\midrule
\multirow[t]{4}{*}{{\bf WD+WD}} & combined $Z$ & $1.2\times 10^{4}$ & $4.0\times 10^{4}$ & 0.96 & $6.4\times 10^{3}$ & $5.7\times 10^{4}$ & 1.86 & $1.8\times 10^{4}$ & $8.8\times 10^{4}$ & 1.62 \\
 & $Z=Z_\odot$ & $1.2\times 10^{4}$ & $4.1\times 10^{4}$ & 1.04 & $6.4\times 10^{3}$ & $5.7\times 10^{4}$ & 1.86 & $1.9\times 10^{4}$ & $9.0\times 10^{4}$ & 1.62 \\
 & $Z=0.1\,Z_\odot$ & $7.3\times 10^{3}$ & $2.2\times 10^{4}$ & 1.14 & $6.4\times 10^{3}$ & $5.7\times 10^{4}$ & 1.86 & $1.4\times 10^{4}$ & $7.6\times 10^{4}$ & 1.72 \\
 & $Z=0.01\,Z_\odot$ & $5.3\times 10^{3}$ & $1.7\times 10^{4}$ & 1.20 & $3.3\times 10^{3}$ & $2.9\times 10^{4}$ & 1.86 & $8.6\times 10^{3}$ & $4.6\times 10^{4}$ & 1.70 \\
\cline{1-11}
\multirow[t]{4}{*}{{\bf O/Ne WD+any}} & combined $Z$ & $4.3\times 10^{3}$ & $2.0\times 10^{4}$ & 1.46 & $9.2\times 10^{3}$ & $4.5\times 10^{4}$ & 1.28 & $1.3\times 10^{4}$ & $6.5\times 10^{4}$ & 1.28 \\
 & $Z=Z_\odot$ & $4.3\times 10^{3}$ & $2.0\times 10^{4}$ & 1.46 & $9.2\times 10^{3}$ & $4.5\times 10^{4}$ & 1.28 & $1.3\times 10^{4}$ & $6.5\times 10^{4}$ & 1.28 \\
 & $Z=0.1\,Z_\odot$ & $4.3\times 10^{3}$ & $2.0\times 10^{4}$ & 1.46 & $9.2\times 10^{3}$ & $4.5\times 10^{4}$ & 1.28 & $1.3\times 10^{4}$ & $6.5\times 10^{4}$ & 1.28 \\
 & $Z=0.01\,Z_\odot$ & $4.3\times 10^{3}$ & $2.0\times 10^{4}$ & 1.46 & $9.2\times 10^{3}$ & $4.5\times 10^{4}$ & 1.28 & $1.3\times 10^{4}$ & $6.5\times 10^{4}$ & 1.28 \\
\cline{1-11}
\multirow[t]{4}{*}{{\bf massive WD+WD}} & combined $Z$ & $5.7\times 10^{3}$ & $2.3\times 10^{4}$ & 0.84 & $3.3\times 10^{3}$ & $3.0\times 10^{4}$ & 1.88 & $9.1\times 10^{3}$ & $4.6\times 10^{4}$ & 1.62 \\
 & $Z=Z_\odot$ & $5.9\times 10^{3}$ & $2.4\times 10^{4}$ & 0.84 & $3.3\times 10^{3}$ & $2.9\times 10^{4}$ & 1.86 & $9.2\times 10^{3}$ & $4.6\times 10^{4}$ & 1.62 \\
 & $Z=0.1\,Z_\odot$ & $4.0\times 10^{3}$ & $1.3\times 10^{4}$ & 0.94 & $3.7\times 10^{3}$ & $3.3\times 10^{4}$ & 1.86 & $7.7\times 10^{3}$ & $4.3\times 10^{4}$ & 1.72 \\
 & $Z=0.01\,Z_\odot$ & $2.5\times 10^{3}$ & $1.4\times 10^{4}$ & 1.50 & $2.4\times 10^{3}$ & $2.1\times 10^{4}$ & 1.86 & $4.9\times 10^{3}$ & $3.4\times 10^{4}$ & 1.70 \\
\cline{1-11}
\bottomrule
\end{tabular}
\end{table*}

\clearpage
\bibliography{references}

@ARTICLE{Shariat25_10k,
       author = {{Shariat}, Cheyanne and {El-Badry}, Kareem and {Naoz}, Smadar},
        title = "{10,000 Resolved Triples from Gaia: Empirical Constraints on Triple Star Populations}",
      journal = {\pasp},
         year = 2025,
        month = sep,
       volume = {137},
       number = {9},
          eid = {094201},
        pages = {094201},
          doi = {10.1088/1538-3873/adfb30},
archivePrefix = {arXiv},
       eprint = {2506.16513},
 primaryClass = {astro-ph.SR},
       adsurl = {https://ui.adsabs.harvard.edu/abs/2025PASP..137i4201S}
}

@ARTICLE{Naoz25,
       author = {{Naoz}, Smadar and {Haiman}, Zolt{\'a}n and {Quataert}, Eliot and {Holzknecht}, Liz},
        title = "{Triples as Links between Binary Black Hole Mergers, Their Electromagnetic Counterparts, and Galactic Black Holes}",
      journal = {\apjl},
         year = 2025,
        month = oct,
       volume = {992},
       number = {1},
          eid = {L12},
        pages = {L12},
          doi = {10.3847/2041-8213/ae0a20},
archivePrefix = {arXiv},
       eprint = {2508.13270},
 primaryClass = {astro-ph.HE},
       adsurl = {https://ui.adsabs.harvard.edu/abs/2025ApJ...992L..12N}
}

@ARTICLE{Lidov1962,
       author = {{Lidov}, M.~L.},
        title = "{The evolution of orbits of artificial satellites of planets under the action of gravitational perturbations of external bodies}",
      journal = {\planss},
         year = 1962,
        month = oct,
       volume = {9},
       number = {10},
        pages = {719-759},
          doi = {10.1016/0032-0633(62)90129-0},
       adsurl = {https://ui.adsabs.harvard.edu/abs/1962P&SS....9..719L}
}

@ARTICLE{Thompson2011,
       author = {{Thompson}, Todd A.},
        title = "{Accelerating Compact Object Mergers in Triple Systems with the Kozai Resonance: A Mechanism for ``Prompt'' Type Ia Supernovae, Gamma-Ray Bursts, and Other Exotica}",
      journal = {\apj},
         year = 2011,
        month = nov,
       volume = {741},
       number = {2},
          eid = {82},
        pages = {82},
          doi = {10.1088/0004-637X/741/2/82},
archivePrefix = {arXiv},
       eprint = {1011.4322},
 primaryClass = {astro-ph.HE},
       adsurl = {https://ui.adsabs.harvard.edu/abs/2011ApJ...741...82T}
}

@ARTICLE{Naoz2016,
       author = {{Naoz}, Smadar},
        title = "{The Eccentric Kozai-Lidov Effect and Its Applications}",
      journal = {\araa},
         year = 2016,
        month = sep,
       volume = {54},
        pages = {441-489},
          doi = {10.1146/annurev-astro-081915-023315},
archivePrefix = {arXiv},
       eprint = {1601.07175},
 primaryClass = {astro-ph.EP},
       adsurl = {https://ui.adsabs.harvard.edu/abs/2016ARA&A..54..441N}
}

@ARTICLE{Naoz2014,
       author = {{Naoz}, Smadar and {Fabrycky}, Daniel C.},
        title = "{Mergers and Obliquities in Stellar Triples}",
      journal = {\apj},
         year = 2014,
        month = oct,
       volume = {793},
       number = {2},
          eid = {137},
        pages = {137},
          doi = {10.1088/0004-637X/793/2/137},
archivePrefix = {arXiv},
       eprint = {1405.5223},
 primaryClass = {astro-ph.SR},
       adsurl = {https://ui.adsabs.harvard.edu/abs/2014ApJ...793..137N}
}

@ARTICLE{SSE,
       author = {{Hurley}, Jarrod R. and {Pols}, Onno R. and {Tout}, Christopher A.},
        title = "{Comprehensive analytic formulae for stellar evolution as a function of mass and metallicity}",
      journal = {\mnras},
         year = 2000,
        month = jul,
       volume = {315},
       number = {3},
        pages = {543-569},
          doi = {10.1046/j.1365-8711.2000.03426.x},
archivePrefix = {arXiv},
       eprint = {astro-ph/0001295},
 primaryClass = {astro-ph},
       adsurl = {https://ui.adsabs.harvard.edu/abs/2000MNRAS.315..543H}
}

@ARTICLE{COSMIC,
       author = {{Breivik}, Katelyn and {Coughlin}, Scott and {Zevin}, Michael and {Rodriguez}, Carl L. and {Kremer}, Kyle and {Ye}, Claire S. and {Andrews}, Jeff J. and {Kurkowski}, Michael and {Digman}, Matthew C. and {Larson}, Shane L. and {Rasio}, Frederic A.},
        title = "{COSMIC Variance in Binary Population Synthesis}",
      journal = {\apj},
         year = 2020,
        month = jul,
       volume = {898},
       number = {1},
          eid = {71},
        pages = {71},
          doi = {10.3847/1538-4357/ab9d85},
archivePrefix = {arXiv},
       eprint = {1911.00903},
 primaryClass = {astro-ph.HE},
       adsurl = {https://ui.adsabs.harvard.edu/abs/2020ApJ...898...71B}
}

@ARTICLE{MA2001,
       author = {{Mardling}, Rosemary A. and {Aarseth}, Sverre J.},
        title = "{Tidal interactions in star cluster simulations}",
      journal = {\mnras},
         year = 2001,
        month = mar,
       volume = {321},
       number = {3},
        pages = {398-420},
          doi = {10.1046/j.1365-8711.2001.03974.x},
       adsurl = {https://ui.adsabs.harvard.edu/abs/2001MNRAS.321..398M}
}

@ARTICLE{Toonen2016,
       author = {{Toonen}, Silvia and {Hamers}, Adrian and {Portegies Zwart}, Simon},
        title = "{The evolution of hierarchical triple star-systems}",
      journal = {Computational Astrophysics and Cosmology},
         year = 2016,
        month = dec,
       volume = {3},
       number = {1},
          eid = {6},
        pages = {6},
          doi = {10.1186/s40668-016-0019-0},
archivePrefix = {arXiv},
       eprint = {1612.06172},
 primaryClass = {astro-ph.SR},
       adsurl = {https://ui.adsabs.harvard.edu/abs/2016ComAC...3....6T}
}

@ARTICLE{Toonen20,
       author = {{Toonen}, S. and {Portegies Zwart}, S. and {Hamers}, A.~S. and {Bandopadhyay}, D.},
        title = "{The evolution of stellar triples. The most common evolutionary pathways}",
      journal = {\aap},
         year = 2020,
        month = aug,
       volume = {640},
          eid = {A16},
        pages = {A16},
          doi = {10.1051/0004-6361/201936835},
archivePrefix = {arXiv},
       eprint = {2004.07848},
 primaryClass = {astro-ph.SR},
       adsurl = {https://ui.adsabs.harvard.edu/abs/2020A&A...640A..16T}
}

@ARTICLE{Shappee13,
       author = {{Shappee}, Benjamin J. and {Thompson}, Todd A.},
        title = "{The Mass-loss-induced Eccentric Kozai Mechanism: A New Channel for the Production of Close Compact Object-Stellar Binaries}",
      journal = {\apj},
         year = 2013,
        month = mar,
       volume = {766},
       number = {1},
          eid = {64},
        pages = {64},
          doi = {10.1088/0004-637X/766/1/64},
archivePrefix = {arXiv},
       eprint = {1204.1053},
 primaryClass = {astro-ph.SR},
       adsurl = {https://ui.adsabs.harvard.edu/abs/2013ApJ...766...64S}
}

@ARTICLE{Stephan16,
       author = {{Stephan}, Alexander P. and {Naoz}, Smadar and {Ghez}, Andrea M. and {Witzel}, Gunther and {Sitarski}, Breann N. and {Do}, Tuan and {Kocsis}, Bence},
        title = "{Merging binaries in the Galactic Center: the eccentric Kozai-Lidov mechanism with stellar evolution}",
      journal = {\mnras},
         year = 2016,
        month = aug,
       volume = {460},
       number = {4},
        pages = {3494-3504},
          doi = {10.1093/mnras/stw1220},
archivePrefix = {arXiv},
       eprint = {1603.02709},
 primaryClass = {astro-ph.SR},
       adsurl = {https://ui.adsabs.harvard.edu/abs/2016MNRAS.460.3494S}
}

@ARTICLE{Angelo22,
       author = {{Angelo}, Isabel and {Naoz}, Smadar and {Petigura}, Erik and {MacDougall}, Mason and {Stephan}, Alexander P. and {Isaacson}, Howard and {Howard}, Andrew W.},
        title = "{Kepler-1656b's Extreme Eccentricity: Signature of a Gentle Giant}",
      journal = {\aj},
         year = 2022,
        month = may,
       volume = {163},
       number = {5},
          eid = {227},
        pages = {227},
          doi = {10.3847/1538-3881/ac6094},
archivePrefix = {arXiv},
       eprint = {2204.00019},
 primaryClass = {astro-ph.EP},
       adsurl = {https://ui.adsabs.harvard.edu/abs/2022AJ....163..227A}
}

@ARTICLE{Hamers13,
       author = {{Hamers}, A.~S. and {Pols}, O.~R. and {Claeys}, J.~S.~W. and {Nelemans}, G.},
        title = "{Population synthesis of triple systems in the context of mergers of carbon-oxygen white dwarfs}",
      journal = {\mnras},
         year = 2013,
        month = apr,
       volume = {430},
       number = {3},
        pages = {2262-2280},
          doi = {10.1093/mnras/stt046},
archivePrefix = {arXiv},
       eprint = {1301.1469},
 primaryClass = {astro-ph.SR},
       adsurl = {https://ui.adsabs.harvard.edu/abs/2013MNRAS.430.2262H}
}

@ARTICLE{Hamers19,
       author = {{Hamers}, Adrian S. and {Thompson}, Todd A.},
        title = "{The Impact of White Dwarf Natal Kicks and Stellar Flybys on the Rates of Type Ia Supernovae in Triple-star Systems}",
      journal = {\apj},
         year = 2019,
        month = sep,
       volume = {882},
       number = {1},
          eid = {24},
        pages = {24},
          doi = {10.3847/1538-4357/ab321f},
archivePrefix = {arXiv},
       eprint = {1904.12881},
 primaryClass = {astro-ph.HE},
       adsurl = {https://ui.adsabs.harvard.edu/abs/2019ApJ...882...24H}
}

@ARTICLE{Kroupa2001,
       author = {{Kroupa}, Pavel and {Tout}, Christopher A. and {Gilmore}, Gerard},
        title = "{The Distribution of Low-Mass Stars in the Galactic Disc}",
      journal = {\mnras},
         year = 1993,
        month = jun,
       volume = {262},
        pages = {545-587},
          doi = {10.1093/mnras/262.3.545},
       adsurl = {https://ui.adsabs.harvard.edu/abs/1993MNRAS.262..545K}
}

@ARTICLE{Katz12,
       author = {{Katz}, Boaz and {Dong}, Subo},
        title = "{The rate of WD-WD head-on collisions may be as high as the SNe Ia rate}",
      journal = {arXiv e-prints},
         year = 2012,
        month = nov,
          eid = {arXiv:1211.4584},
        pages = {arXiv:1211.4584},
          doi = {10.48550/arXiv.1211.4584},
archivePrefix = {arXiv},
       eprint = {1211.4584},
 primaryClass = {astro-ph.SR},
       adsurl = {https://ui.adsabs.harvard.edu/abs/2012arXiv1211.4584K}
}

@ARTICLE{Breivik20,
       author = {{Breivik}, Katelyn and {Mingarelli}, Chiara M.~F. and {Larson}, Shane L.},
        title = "{Constraining Galactic Structure with the LISA White Dwarf Foreground}",
      journal = {\apj},
         year = 2020,
        month = sep,
       volume = {901},
       number = {1},
          eid = {4},
        pages = {4},
          doi = {10.3847/1538-4357/abab99},
archivePrefix = {arXiv},
       eprint = {1912.02200},
 primaryClass = {astro-ph.GA},
       adsurl = {https://ui.adsabs.harvard.edu/abs/2020ApJ...901....4B}
}

@ARTICLE{Stegmann22,
       author = {{Stegmann}, Jakob and {Antonini}, Fabio and {Moe}, Maxwell},
        title = "{Evolution of massive stellar triples and implications for compact object binary formation}",
      journal = {\mnras},
         year = 2022,
        month = oct,
       volume = {516},
       number = {1},
        pages = {1406-1427},
          doi = {10.1093/mnras/stac2192},
archivePrefix = {arXiv},
       eprint = {2112.10786},
 primaryClass = {astro-ph.SR},
       adsurl = {https://ui.adsabs.harvard.edu/abs/2022MNRAS.516.1406S}
}

@ARTICLE{Liu23Review,
       author = {{Liu}, Zheng-Wei and {Roepke}, Friedrich K. and {Han}, Zhanwen},
        title = "{Type Ia Supernova Explosions in Binary Systems: A Review}",
      journal = {arXiv e-prints},
         year = 2023,
        month = may,
          eid = {arXiv:2305.13305},
        pages = {arXiv:2305.13305},
          doi = {10.48550/arXiv.2305.13305},
archivePrefix = {arXiv},
       eprint = {2305.13305},
 primaryClass = {astro-ph.HE},
       adsurl = {https://ui.adsabs.harvard.edu/abs/2023arXiv230513305L}
}

@ARTICLE{Tokovinin14,
       author = {{Tokovinin}, Andrei},
        title = "{From Binaries to Multiples. II. Hierarchical Multiplicity of F and G Dwarfs}",
      journal = {\aj},
         year = 2014,
        month = apr,
       volume = {147},
       number = {4},
          eid = {87},
        pages = {87},
          doi = {10.1088/0004-6256/147/4/87},
archivePrefix = {arXiv},
       eprint = {1401.6827},
 primaryClass = {astro-ph.SR},
       adsurl = {https://ui.adsabs.harvard.edu/abs/2014AJ....147...87T}
}

@INPROCEEDINGS{Offner23,
       author = {{Offner}, S.~S.~R. and {Moe}, M. and {Kratter}, K.~M. and {Sadavoy}, S.~I. and {Jensen}, E.~L.~N. and {Tobin}, J.~J.},
        title = "{The Origin and Evolution of Multiple Star Systems}",
    booktitle = {Protostars and Planets VII},
         year = 2023,
       editor = {{Inutsuka}, S. and {Aikawa}, Y. and {Muto}, T. and {Tomida}, K. and {Tamura}, M.},
       series = {Astronomical Society of the Pacific Conference Series},
       volume = {534},
        month = jul,
        pages = {275},
          doi = {10.48550/arXiv.2203.10066},
archivePrefix = {arXiv},
       eprint = {2203.10066},
 primaryClass = {astro-ph.SR},
       adsurl = {https://ui.adsabs.harvard.edu/abs/2023ASPC..534..275O}
}

@ARTICLE{Tokovinin14b,
       author = {{Tokovinin}, Andrei},
        title = "{From Binaries to Multiples. II. Hierarchical Multiplicity of F and G Dwarfs}",
      journal = {\aj},
         year = 2014,
        month = apr,
       volume = {147},
       number = {4},
          eid = {87},
        pages = {87},
          doi = {10.1088/0004-6256/147/4/87},
archivePrefix = {arXiv},
       eprint = {1401.6827},
 primaryClass = {astro-ph.SR},
       adsurl = {https://ui.adsabs.harvard.edu/abs/2014AJ....147...87T}
}

@ARTICLE{Knigge22,
       author = {{Knigge}, C. and {Toonen}, S. and {Boekholt}, T.~C.~N.},
        title = "{A triple star origin for T Pyx and other short-period recurrent novae}",
      journal = {\mnras},
         year = 2022,
        month = aug,
       volume = {514},
       number = {2},
        pages = {1895-1907},
          doi = {10.1093/mnras/stac1336},
archivePrefix = {arXiv},
       eprint = {2205.00014},
 primaryClass = {astro-ph.SR},
       adsurl = {https://ui.adsabs.harvard.edu/abs/2022MNRAS.514.1895K}
}

@ARTICLE{Shariat25Merge,
       author = {{Shariat}, Cheyanne and {Naoz}, Smadar and {El-Badry}, Kareem and {Rodriguez}, Antonio C. and {Hansen}, Bradley M.~S. and {Angelo}, Isabel and {Stephan}, Alexander P.},
        title = "{Once a Triple, Not Always a Triple: The Evolution of Hierarchical Triples That Yield Merged Inner Binaries}",
      journal = {\apj},
         year = 2025,
        month = jan,
       volume = {978},
       number = {1},
          eid = {47},
        pages = {47},
          doi = {10.3847/1538-4357/ad944a},
archivePrefix = {arXiv},
       eprint = {2407.06257},
 primaryClass = {astro-ph.SR},
       adsurl = {https://ui.adsabs.harvard.edu/abs/2025ApJ...978...47S}
}

@ARTICLE{NaozLMXB,
       author = {{Naoz}, Smadar and {Fragos}, Tassos and {Geller}, Aaron and {Stephan}, Alexander P. and {Rasio}, Frederic A.},
        title = "{Formation of Black Hole Low-mass X-Ray Binaries in Hierarchical Triple Systems}",
      journal = {\apjl},
         year = 2016,
        month = may,
       volume = {822},
       number = {2},
          eid = {L24},
        pages = {L24},
          doi = {10.3847/2041-8205/822/2/L24},
archivePrefix = {arXiv},
       eprint = {1510.02093},
 primaryClass = {astro-ph.HE},
       adsurl = {https://ui.adsabs.harvard.edu/abs/2016ApJ...822L..24N}
}

@ARTICLE{Tokovinin22_resolvedtriples,
       author = {{Tokovinin}, Andrei},
        title = "{Resolved Gaia Triples}",
      journal = {\apj},
         year = 2022,
        month = feb,
       volume = {926},
       number = {1},
          eid = {1},
        pages = {1},
          doi = {10.3847/1538-4357/ac4584},
archivePrefix = {arXiv},
       eprint = {2112.11943},
 primaryClass = {astro-ph.SR},
       adsurl = {https://ui.adsabs.harvard.edu/abs/2022ApJ...926....1T}
}

@ARTICLE{Raghavan2010,
       author = {{Raghavan}, Deepak and {McAlister}, Harold A. and {Henry}, Todd J. and {Latham}, David W. and {Marcy}, Geoffrey W. and {Mason}, Brian D. and {Gies}, Douglas R. and {White}, Russel J. and {ten Brummelaar}, Theo A.},
        title = "{A Survey of Stellar Families: Multiplicity of Solar-type Stars}",
      journal = {\apjs},
         year = 2010,
        month = sep,
       volume = {190},
       number = {1},
        pages = {1-42},
          doi = {10.1088/0067-0049/190/1/1},
archivePrefix = {arXiv},
       eprint = {1007.0414},
 primaryClass = {astro-ph.SR},
       adsurl = {https://ui.adsabs.harvard.edu/abs/2010ApJS..190....1R}
}

@ARTICLE{Kozai1962,
       author = {{Kozai}, Yoshihide},
        title = "{Secular perturbations of asteroids with high inclination and eccentricity}",
      journal = {\aj},
         year = 1962,
        month = nov,
       volume = {67},
        pages = {591-598},
          doi = {10.1086/108790},
       adsurl = {https://ui.adsabs.harvard.edu/abs/1962AJ.....67..591K}
}

@ARTICLE{Michaely2020,
       author = {{Michaely}, Erez and {Perets}, Hagai B.},
        title = "{High rate of gravitational waves mergers from flyby perturbations of wide black hole triples in the field}",
      journal = {\mnras},
         year = 2020,
        month = nov,
       volume = {498},
       number = {4},
        pages = {4924-4935},
          doi = {10.1093/mnras/staa2720},
archivePrefix = {arXiv},
       eprint = {2008.01094},
 primaryClass = {astro-ph.HE},
       adsurl = {https://ui.adsabs.harvard.edu/abs/2020MNRAS.498.4924M}
}

@article{BSE,
    author = {Hurley, Jarrod R. and Tout, Christopher A. and Pols, Onno R.},
    title = "{Evolution of binary stars and the effect of tides on binary populations}",
    journal = {Monthly Notices of the Royal Astronomical Society},
    volume = {329},
    number = {4},
    pages = {897-928},
    year = {2002},
    month = {02},
    issn = {0035-8711},
    doi = {10.1046/j.1365-8711.2002.05038.x},
    url = {https://doi.org/10.1046/j.1365-8711.2002.05038.x},
    eprint = {https://academic.oup.com/mnras/article-pdf/329/4/897/18418535/329-4-897.pdf},
}

@ARTICLE{Kaib2014,
       author = {{Kaib}, Nathan A. and {Raymond}, Sean N.},
        title = "{Very Wide Binary Stars as the Primary Source of Stellar Collisions in the Galaxy}",
      journal = {\apj},
         year = 2014,
        month = feb,
       volume = {782},
       number = {2},
          eid = {60},
        pages = {60},
          doi = {10.1088/0004-637X/782/2/60},
archivePrefix = {arXiv},
       eprint = {1309.3272},
 primaryClass = {astro-ph.SR},
       adsurl = {https://ui.adsabs.harvard.edu/abs/2014ApJ...782...60K}
}

@ARTICLE{Naoz2013sec,
       author = {{Naoz}, Smadar and {Farr}, Will M. and {Lithwick}, Yoram and {Rasio}, Frederic A. and {Teyssandier}, Jean},
        title = "{Secular dynamics in hierarchical three-body systems}",
      journal = {\mnras},
         year = 2013,
        month = may,
       volume = {431},
       number = {3},
        pages = {2155-2171},
          doi = {10.1093/mnras/stt302},
archivePrefix = {arXiv},
       eprint = {1107.2414},
 primaryClass = {astro-ph.EP},
       adsurl = {https://ui.adsabs.harvard.edu/abs/2013MNRAS.431.2155N}
}

@ARTICLE{Naoz2013GR,
       author = {{Naoz}, Smadar and {Kocsis}, Bence and {Loeb}, Abraham and {Yunes}, Nicol{\'a}s},
        title = "{Resonant Post-Newtonian Eccentricity Excitation in Hierarchical Three-body Systems}",
      journal = {\apj},
         year = 2013,
        month = aug,
       volume = {773},
       number = {2},
          eid = {187},
        pages = {187},
          doi = {10.1088/0004-637X/773/2/187},
archivePrefix = {arXiv},
       eprint = {1206.4316},
 primaryClass = {astro-ph.SR},
       adsurl = {https://ui.adsabs.harvard.edu/abs/2013ApJ...773..187N}
}

@ARTICLE{Hut80,
       author = {{Hut}, P.},
        title = "{Stability of tidal equilibrium}",
      journal = {\aap},
         year = 1980,
        month = dec,
       volume = {92},
       number = {1-2},
        pages = {167-170},
       adsurl = {https://ui.adsabs.harvard.edu/abs/1980A&A....92..167H}
}

@ARTICLE{Eggleton98,
       author = {{Eggleton}, Peter P. and {Kiseleva}, Ludmila G. and {Hut}, Piet},
        title = "{The Equilibrium Tide Model for Tidal Friction}",
      journal = {\apj},
         year = 1998,
        month = may,
       volume = {499},
       number = {2},
        pages = {853-870},
          doi = {10.1086/305670},
archivePrefix = {arXiv},
       eprint = {astro-ph/9801246},
 primaryClass = {astro-ph},
       adsurl = {https://ui.adsabs.harvard.edu/abs/1998ApJ...499..853E}
}

@ARTICLE{Zahn77,
       author = {{Zahn}, J. -P.},
        title = "{Tidal friction in close binary systems.}",
      journal = {\aap},
         year = 1977,
        month = may,
       volume = {57},
        pages = {383-394},
       adsurl = {https://ui.adsabs.harvard.edu/abs/1977A&A....57..383Z}
}

@ARTICLE{Moe17,
       author = {{Moe}, Maxwell and {Di Stefano}, Rosanne},
        title = "{Mind Your Ps and Qs: The Interrelation between Period (P) and Mass-ratio (Q) Distributions of Binary Stars}",
      journal = {\apjs},
         year = 2017,
        month = jun,
       volume = {230},
       number = {2},
          eid = {15},
        pages = {15},
          doi = {10.3847/1538-4365/aa6fb6},
archivePrefix = {arXiv},
       eprint = {1606.05347},
 primaryClass = {astro-ph.SR},
       adsurl = {https://ui.adsabs.harvard.edu/abs/2017ApJS..230...15M}
}

@ARTICLE{Toonen18,
       author = {{Toonen}, S. and {Perets}, H.~B. and {Hamers}, A.~S.},
        title = "{Rate of WD-WD head-on collisions in isolated triples is too low to explain standard type Ia supernovae}",
      journal = {\aap},
         year = 2018,
        month = feb,
       volume = {610},
          eid = {A22},
        pages = {A22},
          doi = {10.1051/0004-6361/201731874},
archivePrefix = {arXiv},
       eprint = {1709.00422},
 primaryClass = {astro-ph.HE},
       adsurl = {https://ui.adsabs.harvard.edu/abs/2018A&A...610A..22T}
}

@ARTICLE{Shariat23,
       author = {{Shariat}, Cheyanne and {Naoz}, Smadar and {Hansen}, Bradley M.~S. and {Angelo}, Isabel and {Michaely}, Erez and {Stephan}, Alexander P.},
        title = "{Dynamical Evolution of White Dwarfs in Triples in the Era of Gaia}",
      journal = {\apjl},
         year = 2023,
        month = sep,
       volume = {955},
       number = {1},
          eid = {L14},
        pages = {L14},
          doi = {10.3847/2041-8213/acf76b},
archivePrefix = {arXiv},
       eprint = {2306.13130},
 primaryClass = {astro-ph.SR},
       adsurl = {https://ui.adsabs.harvard.edu/abs/2023ApJ...955L..14S}
}

@ARTICLE{Shariat25CV,
       author = {{Shariat}, Cheyanne and {El-Badry}, Kareem and {Naoz}, Smadar and {Rodriguez}, Antonio C. and {van Roestel}, Jan},
        title = "{Cataclysmic Variables in Triples: Formation Models and New Discoveries}",
      journal = {\pasp},
         year = 2025,
        month = jul,
       volume = {137},
       number = {7},
          eid = {074201},
        pages = {074201},
          doi = {10.1088/1538-3873/add5a1},
archivePrefix = {arXiv},
       eprint = {2501.14025},
 primaryClass = {astro-ph.SR},
       adsurl = {https://ui.adsabs.harvard.edu/abs/2025PASP..137g4201S}
}

@ARTICLE{Shariat24LMXB,
       author = {{Shariat}, Cheyanne and {Naoz}, Smadar and {El-Badry}, Kareem and {Rocha}, Kyle Akira and {Kalogera}, Vicky and {Stephan}, Alexander P. and {Burdge}, Kevin B. and {Angelo}, Isabel},
        title = "{Triple Evolution Pathways to Black Hole Low-mass X-Ray Binaries: Insights from V404 Cygni}",
      journal = {\apj},
         year = 2025,
        month = apr,
       volume = {983},
       number = {2},
          eid = {115},
        pages = {115},
          doi = {10.3847/1538-4357/adbf01},
       adsurl = {https://ui.adsabs.harvard.edu/abs/2025ApJ...983..115S}
}

@ARTICLE{Perets25,
       author = {{Perets}, Hagai B.},
        title = "{Evolution of Triple Stars}",
      journal = {arXiv e-prints},
         year = 2025,
        month = apr,
          eid = {arXiv:2504.02939},
        pages = {arXiv:2504.02939},
          doi = {10.48550/arXiv.2504.02939},
archivePrefix = {arXiv},
       eprint = {2504.02939},
 primaryClass = {astro-ph.SR},
       adsurl = {https://ui.adsabs.harvard.edu/abs/2025arXiv250402939P}
}

@ARTICLE{MoeKratter21,
       author = {{Moe}, Maxwell and {Kratter}, Kaitlin M.},
        title = "{Impact of binary stars on planet statistics - I. Planet occurrence rates and trends with stellar mass}",
      journal = {\mnras},
         year = 2021,
        month = nov,
       volume = {507},
       number = {3},
        pages = {3593-3611},
          doi = {10.1093/mnras/stab2328},
archivePrefix = {arXiv},
       eprint = {1912.01699},
 primaryClass = {astro-ph.EP},
       adsurl = {https://ui.adsabs.harvard.edu/abs/2021MNRAS.507.3593M}
}

@ARTICLE{Maoz12_review,
       author = {{Maoz}, D. and {Mannucci}, F.},
        title = "{Type-Ia Supernova Rates and the Progenitor Problem: A Review}",
      journal = {\pasa},
         year = 2012,
        month = jan,
       volume = {29},
       number = {4},
        pages = {447-465},
          doi = {10.1071/AS11052},
archivePrefix = {arXiv},
       eprint = {1111.4492},
 primaryClass = {astro-ph.CO},
       adsurl = {https://ui.adsabs.harvard.edu/abs/2012PASA...29..447M}
}

@ARTICLE{Maoz12,
       author = {{Maoz}, Dan and {Mannucci}, Filippo and {Brandt}, Timothy D.},
        title = "{The delay-time distribution of Type Ia supernovae from Sloan II}",
      journal = {\mnras},
         year = 2012,
        month = nov,
       volume = {426},
       number = {4},
        pages = {3282-3294},
          doi = {10.1111/j.1365-2966.2012.21871.x},
archivePrefix = {arXiv},
       eprint = {1206.0465},
 primaryClass = {astro-ph.CO},
       adsurl = {https://ui.adsabs.harvard.edu/abs/2012MNRAS.426.3282M}
}

@ARTICLE{Lorimer07,
       author = {{Lorimer}, D.~R. and {Bailes}, M. and {McLaughlin}, M.~A. and {Narkevic}, D.~J. and {Crawford}, F.},
        title = "{A Bright Millisecond Radio Burst of Extragalactic Origin}",
      journal = {Science},
         year = 2007,
        month = nov,
       volume = {318},
       number = {5851},
        pages = {777},
          doi = {10.1126/science.1147532},
archivePrefix = {arXiv},
       eprint = {0709.4301},
 primaryClass = {astro-ph},
       adsurl = {https://ui.adsabs.harvard.edu/abs/2007Sci...318..777L}
}

@ARTICLE{Thornton13,
       author = {{Thornton}, D. and {Stappers}, B. and {Bailes}, M. and {Barsdell}, B. and {Bates}, S. and {Bhat}, N.~D.~R. and {Burgay}, M. and {Burke-Spolaor}, S. and {Champion}, D.~J. and {Coster}, P. and {D'Amico}, N. and {Jameson}, A. and {Johnston}, S. and {Keith}, M. and {Kramer}, M. and {Levin}, L. and {Milia}, S. and {Ng}, C. and {Possenti}, A. and {van Straten}, W.},
        title = "{A Population of Fast Radio Bursts at Cosmological Distances}",
      journal = {Science},
         year = 2013,
        month = jul,
       volume = {341},
       number = {6141},
        pages = {53-56},
          doi = {10.1126/science.1236789},
archivePrefix = {arXiv},
       eprint = {1307.1628},
 primaryClass = {astro-ph.HE},
       adsurl = {https://ui.adsabs.harvard.edu/abs/2013Sci...341...53T}
}

@ARTICLE{Petroff19,
       author = {{Petroff}, E. and {Hessels}, J.~W.~T. and {Lorimer}, D.~R.},
        title = "{Fast radio bursts}",
      journal = {\aapr},
         year = 2019,
        month = dec,
       volume = {27},
       number = {1},
          eid = {4},
        pages = {4},
          doi = {10.1007/s00159-019-0116-6},
archivePrefix = {arXiv},
       eprint = {1904.07947},
 primaryClass = {astro-ph.HE},
       adsurl = {https://ui.adsabs.harvard.edu/abs/2019A&ARv..27....4P}
}

@ARTICLE{Bochenek20,
       author = {{Bochenek}, C.~D. and {Ravi}, V. and {Belov}, K.~V. and {Hallinan}, G. and {Kocz}, J. and {Kulkarni}, S.~R. and {McKenna}, D.~L.},
        title = "{A fast radio burst associated with a Galactic magnetar}",
      journal = {\nat},
         year = 2020,
        month = nov,
       volume = {587},
       number = {7832},
        pages = {59-62},
          doi = {10.1038/s41586-020-2872-x},
archivePrefix = {arXiv},
       eprint = {2005.10828},
 primaryClass = {astro-ph.HE},
       adsurl = {https://ui.adsabs.harvard.edu/abs/2020Natur.587...59B}
}

@ARTICLE{Totani13,
       author = {{Totani}, Tomonori},
        title = "{Cosmological Fast Radio Bursts from Binary Neutron Star Mergers}",
      journal = {\pasj},
         year = 2013,
        month = oct,
       volume = {65},
       number = {5},
          eid = {L12},
        pages = {L12},
          doi = {10.1093/pasj/65.5.L12},
archivePrefix = {arXiv},
       eprint = {1307.4985},
 primaryClass = {astro-ph.HE},
       adsurl = {https://ui.adsabs.harvard.edu/abs/2013PASJ...65L..12T}
}

@ARTICLE{Wang16,
       author = {{Wang}, Jie-Shuang and {Yang}, Yuan-Pei and {Wu}, Xue-Feng and {Dai}, Zi-Gao and {Wang}, Fa-Yin},
        title = "{Fast Radio Bursts from the Inspiral of Double Neutron Stars}",
      journal = {\apjl},
         year = 2016,
        month = may,
       volume = {822},
       number = {1},
          eid = {L7},
        pages = {L7},
          doi = {10.3847/2041-8205/822/1/L7},
archivePrefix = {arXiv},
       eprint = {1603.02014},
 primaryClass = {astro-ph.HE},
       adsurl = {https://ui.adsabs.harvard.edu/abs/2016ApJ...822L...7W}
}

@ARTICLE{Nomoto91,
       author = {{Nomoto}, Ken'ichi and {Kondo}, Yoji},
        title = "{Conditions for Accretion-induced Collapse of White Dwarfs}",
      journal = {\apjl},
         year = 1991,
        month = jan,
       volume = {367},
        pages = {L19},
          doi = {10.1086/185922},
       adsurl = {https://ui.adsabs.harvard.edu/abs/1991ApJ...367L..19N}
}

@ARTICLE{Schwab21,
       author = {{Schwab}, Josiah},
        title = "{Evolutionary Models for the Remnant of the Merger of Two Carbon-Oxygen Core White Dwarfs}",
      journal = {\apj},
         year = 2021,
        month = jan,
       volume = {906},
       number = {1},
          eid = {53},
        pages = {53},
          doi = {10.3847/1538-4357/abc87e},
archivePrefix = {arXiv},
       eprint = {2011.03546},
 primaryClass = {astro-ph.SR},
       adsurl = {https://ui.adsabs.harvard.edu/abs/2021ApJ...906...53S}
}

@ARTICLE{Belczynski08,
       author = {{Belczynski}, Krzysztof and {Kalogera}, Vassiliki and {Rasio}, Frederic A. and {Taam}, Ronald E. and {Zezas}, Andreas and {Bulik}, Tomasz and {Maccarone}, Thomas J. and {Ivanova}, Natalia},
        title = "{Compact Object Modeling with the StarTrack Population Synthesis Code}",
      journal = {\apjs},
         year = 2008,
        month = jan,
       volume = {174},
       number = {1},
        pages = {223-260},
          doi = {10.1086/521026},
archivePrefix = {arXiv},
       eprint = {astro-ph/0511811},
 primaryClass = {astro-ph},
       adsurl = {https://ui.adsabs.harvard.edu/abs/2008ApJS..174..223B}
}

@ARTICLE{Madau14,
       author = {{Madau}, Piero and {Dickinson}, Mark},
        title = "{Cosmic Star-Formation History}",
      journal = {\araa},
         year = 2014,
        month = aug,
       volume = {52},
        pages = {415-486},
          doi = {10.1146/annurev-astro-081811-125615},
archivePrefix = {arXiv},
       eprint = {1403.0007},
 primaryClass = {astro-ph.CO},
       adsurl = {https://ui.adsabs.harvard.edu/abs/2014ARA&A..52..415M}
}

@ARTICLE{Heintz20,
       author = {{Heintz}, Kasper E. and {Prochaska}, J. Xavier and {Simha}, Sunil and {Platts}, Emma and {Fong}, Wen-fai and {Tejos}, Nicolas and {Ryder}, Stuart D. and {Aggerwal}, Kshitij and {Bhandari}, Shivani and {Day}, Cherie K. and {Deller}, Adam T. and {Kilpatrick}, Charles D. and {Law}, Casey J. and {Macquart}, Jean-Pierre and {Mannings}, Alexandra and {Marnoch}, Lachlan J. and {Sadler}, Elaine M. and {Shannon}, Ryan M.},
        title = "{Host Galaxy Properties and Offset Distributions of Fast Radio Bursts: Implications for Their Progenitors}",
      journal = {\apj},
         year = 2020,
        month = nov,
       volume = {903},
       number = {2},
          eid = {152},
        pages = {152},
          doi = {10.3847/1538-4357/abb6fb},
archivePrefix = {arXiv},
       eprint = {2009.10747},
 primaryClass = {astro-ph.GA},
       adsurl = {https://ui.adsabs.harvard.edu/abs/2020ApJ...903..152H}
}

@ARTICLE{Yamasaki25,
       author = {{Yamasaki}, Shotaro and {Hashimoto}, Tetsuya and {Kusakabe}, Haruka and {Goto}, Tomotsugu},
        title = "{No Metallicity Preference in Fast Radio Burst Host Galaxies}",
      journal = {arXiv e-prints},
         year = 2025,
        month = aug,
          eid = {arXiv:2508.07688},
        pages = {arXiv:2508.07688},
          doi = {10.48550/arXiv.2508.07688},
archivePrefix = {arXiv},
       eprint = {2508.07688},
 primaryClass = {astro-ph.HE},
       adsurl = {https://ui.adsabs.harvard.edu/abs/2025arXiv250807688Y}
}

@ARTICLE{Law24,
       author = {{Law}, Casey J. and {Sharma}, Kritti and {Ravi}, Vikram and {Chen}, Ge and {Catha}, Morgan and {Connor}, Liam and {Faber}, Jakob T. and {Hallinan}, Gregg and {Harnach}, Charlie and {Hellbourg}, Greg and {Hobbs}, Rick and {Hodge}, David and {Hodges}, Mark and {Lamb}, James W. and {Rasmussen}, Paul and {Sherman}, Myles B. and {Shi}, Jun and {Simard}, Dana and {Squillace}, Reynier and {Weinreb}, Sander and {Woody}, David P. and {Yurk}, Nitika Yadlapalli},
        title = "{Deep Synoptic Array Science: First FRB and Host Galaxy Catalog}",
      journal = {\apj},
         year = 2024,
        month = may,
       volume = {967},
       number = {1},
          eid = {29},
        pages = {29},
          doi = {10.3847/1538-4357/ad3736},
archivePrefix = {arXiv},
       eprint = {2307.03344},
 primaryClass = {astro-ph.HE},
       adsurl = {https://ui.adsabs.harvard.edu/abs/2024ApJ...967...29L}
}

@ARTICLE{Platts19,
       author = {{Platts}, E. and {Weltman}, A. and {Walters}, A. and {Tendulkar}, S.~P. and {Gordin}, J.~E.~B. and {Kandhai}, S.},
        title = "{A living theory catalogue for fast radio bursts}",
      journal = {\physrep},
         year = 2019,
        month = aug,
       volume = {821},
        pages = {1-27},
          doi = {10.1016/j.physrep.2019.06.003},
archivePrefix = {arXiv},
       eprint = {1810.05836},
 primaryClass = {astro-ph.HE},
       adsurl = {https://ui.adsabs.harvard.edu/abs/2019PhR...821....1P}
}

@ARTICLE{CHIME20,
       author = {{CHIME/FRB Collaboration} and {Andersen}, B.~C. and {Bandura}, K.~M. and {Bhardwaj}, M. and {Bij}, A. and {Boyce}, M.~M. and {Boyle}, P.~J. and {Brar}, C. and {Cassanelli}, T. and {Chawla}, P. and {Chen}, T. and {Cliche}, J. -F. and {Cook}, A. and {Cubranic}, D. and {Curtin}, A.~P. and {Denman}, N.~T. and {Dobbs}, M. and {Dong}, F.~Q. and {Fandino}, M. and {Fonseca}, E. and {Gaensler}, B.~M. and {Giri}, U. and {Good}, D.~C. and {Halpern}, M. and {Hill}, A.~S. and {Hinshaw}, G.~F. and {H{\"o}fer}, C. and {Josephy}, A. and {Kania}, J.~W. and {Kaspi}, V.~M. and {Landecker}, T.~L. and {Leung}, C. and {Li}, D.~Z. and {Lin}, H. -H. and {Masui}, K.~W. and {McKinven}, R. and {Mena-Parra}, J. and {Merryfield}, M. and {Meyers}, B.~W. and {Michilli}, D. and {Milutinovic}, N. and {Mirhosseini}, A. and {M{\"u}nchmeyer}, M. and {Naidu}, A. and {Newburgh}, L.~B. and {Ng}, C. and {Patel}, C. and {Pen}, U. -L. and {Pinsonneault-Marotte}, T. and {Pleunis}, Z. and {Quine}, B.~M. and {Rafiei-Ravandi}, M. and {Rahman}, M. and {Ransom}, S.~M. and {Renard}, A. and {Sanghavi}, P. and {Scholz}, P. and {Shaw}, J.~R. and {Shin}, K. and {Siegel}, S.~R. and {Singh}, S. and {Smegal}, R.~J. and {Smith}, K.~M. and {Stairs}, I.~H. and {Tan}, C.~M. and {Tendulkar}, S.~P. and {Tretyakov}, I. and {Vanderlinde}, K. and {Wang}, H. and {Wulf}, D. and {Zwaniga}, A.~V.},
        title = "{A bright millisecond-duration radio burst from a Galactic magnetar}",
      journal = {\nat},
         year = 2020,
        month = nov,
       volume = {587},
       number = {7832},
        pages = {54-58},
          doi = {10.1038/s41586-020-2863-y},
archivePrefix = {arXiv},
       eprint = {2005.10324},
 primaryClass = {astro-ph.HE},
       adsurl = {https://ui.adsabs.harvard.edu/abs/2020Natur.587...54C}
}

@ARTICLE{CHIME21,
       author = {{CHIME/FRB Collaboration} and {Amiri}, Mandana and {Andersen}, Bridget C. and {Bandura}, Kevin and {Berger}, Sabrina and {Bhardwaj}, Mohit and {Boyce}, Michelle M. and {Boyle}, P.~J. and {Brar}, Charanjot and {Breitman}, Daniela and {Cassanelli}, Tomas and {Chawla}, Pragya and {Chen}, Tianyue and {Cliche}, J. -F. and {Cook}, Amanda and {Cubranic}, Davor and {Curtin}, Alice P. and {Deng}, Meiling and {Dobbs}, Matt and {Dong}, Fengqiu Adam and {Eadie}, Gwendolyn and {Fandino}, Mateus and {Fonseca}, Emmanuel and {Gaensler}, B.~M. and {Giri}, Utkarsh and {Good}, Deborah C. and {Halpern}, Mark and {Hill}, Alex S. and {Hinshaw}, Gary and {Josephy}, Alexander and {Kaczmarek}, Jane F. and {Kader}, Zarif and {Kania}, Joseph W. and {Kaspi}, Victoria M. and {Landecker}, T.~L. and {Lang}, Dustin and {Leung}, Calvin and {Li}, Dongzi and {Lin}, Hsiu-Hsien and {Masui}, Kiyoshi W. and {McKinven}, Ryan and {Mena-Parra}, Juan and {Merryfield}, Marcus and {Meyers}, Bradley W. and {Michilli}, Daniele and {Milutinovic}, Nikola and {Mirhosseini}, Arash and {M{\"u}nchmeyer}, Moritz and {Naidu}, Arun and {Newburgh}, Laura and {Ng}, Cherry and {Patel}, Chitrang and {Pen}, Ue-Li and {Petroff}, Emily and {Pinsonneault-Marotte}, Tristan and {Pleunis}, Ziggy and {Rafiei-Ravandi}, Masoud and {Rahman}, Mubdi and {Ransom}, Scott M. and {Renard}, Andre and {Sanghavi}, Pranav and {Scholz}, Paul and {Shaw}, J. Richard and {Shin}, Kaitlyn and {Siegel}, Seth R. and {Sikora}, Andrew E. and {Singh}, Saurabh and {Smith}, Kendrick M. and {Stairs}, Ingrid and {Tan}, Chia Min and {Tendulkar}, S.~P. and {Vanderlinde}, Keith and {Wang}, Haochen and {Wulf}, Dallas and {Zwaniga}, A.~V.},
        title = "{The First CHIME/FRB Fast Radio Burst Catalog}",
      journal = {\apjs},
         year = 2021,
        month = dec,
       volume = {257},
       number = {2},
          eid = {59},
        pages = {59},
          doi = {10.3847/1538-4365/ac33ab},
archivePrefix = {arXiv},
       eprint = {2106.04352},
 primaryClass = {astro-ph.HE},
       adsurl = {https://ui.adsabs.harvard.edu/abs/2021ApJS..257...59C}
}

@ARTICLE{Bannister19,
       author = {{Bannister}, K.~W. and {Deller}, A.~T. and {Phillips}, C. and {Macquart}, J. -P. and {Prochaska}, J.~X. and {Tejos}, N. and {Ryder}, S.~D. and {Sadler}, E.~M. and {Shannon}, R.~M. and {Simha}, S. and {Day}, C.~K. and {McQuinn}, M. and {North-Hickey}, F.~O. and {Bhandari}, S. and {Arcus}, W.~R. and {Bennert}, V.~N. and {Burchett}, J. and {Bouwhuis}, M. and {Dodson}, R. and {Ekers}, R.~D. and {Farah}, W. and {Flynn}, C. and {James}, C.~W. and {Kerr}, M. and {Lenc}, E. and {Mahony}, E.~K. and {O'Meara}, J. and {Os{\l}owski}, S. and {Qiu}, H. and {Treu}, T. and {U}, V. and {Bateman}, T.~J. and {Bock}, D.~C. -J. and {Bolton}, R.~J. and {Brown}, A. and {Bunton}, J.~D. and {Chippendale}, A.~P. and {Cooray}, F.~R. and {Cornwell}, T. and {Gupta}, N. and {Hayman}, D.~B. and {Kesteven}, M. and {Koribalski}, B.~S. and {MacLeod}, A. and {McClure-Griffiths}, N.~M. and {Neuhold}, S. and {Norris}, R.~P. and {Pilawa}, M.~A. and {Qiao}, R. -Y. and {Reynolds}, J. and {Roxby}, D.~N. and {Shimwell}, T.~W. and {Voronkov}, M.~A. and {Wilson}, C.~D.},
        title = "{A single fast radio burst localized to a massive galaxy at cosmological distance}",
      journal = {Science},
         year = 2019,
        month = aug,
       volume = {365},
       number = {6453},
        pages = {565-570},
          doi = {10.1126/science.aaw5903},
archivePrefix = {arXiv},
       eprint = {1906.11476},
 primaryClass = {astro-ph.HE},
       adsurl = {https://ui.adsabs.harvard.edu/abs/2019Sci...365..565B}
}

@ARTICLE{Prochaska19,
       author = {{Prochaska}, J. Xavier and {Macquart}, Jean-Pierre and {McQuinn}, Matthew and {Simha}, Sunil and {Shannon}, Ryan M. and {Day}, Cherie K. and {Marnoch}, Lachlan and {Ryder}, Stuart and {Deller}, Adam and {Bannister}, Keith W. and {Bhandari}, Shivani and {Bordoloi}, Rongmon and {Bunton}, John and {Cho}, Hyerin and {Flynn}, Chris and {Mahony}, Elizabeth K. and {Phillips}, Chris and {Qiu}, Hao and {Tejos}, Nicolas},
        title = "{The low density and magnetization of a massive galaxy halo exposed by a fast radio burst}",
      journal = {Science},
         year = 2019,
        month = oct,
       volume = {366},
       number = {6462},
        pages = {231-234},
          doi = {10.1126/science.aay0073},
archivePrefix = {arXiv},
       eprint = {1909.11681},
 primaryClass = {astro-ph.GA},
       adsurl = {https://ui.adsabs.harvard.edu/abs/2019Sci...366..231P}
}

@ARTICLE{Bhandari20,
       author = {{Bhandari}, Shivani and {Sadler}, Elaine M. and {Prochaska}, J. Xavier and {Simha}, Sunil and {Ryder}, Stuart D. and {Marnoch}, Lachlan and {Bannister}, Keith W. and {Macquart}, Jean-Pierre and {Flynn}, Chris and {Shannon}, Ryan M. and {Tejos}, Nicolas and {Corro-Guerra}, Felipe and {Day}, Cherie K. and {Deller}, Adam T. and {Ekers}, Ron and {Lopez}, Sebastian and {Mahony}, Elizabeth K. and {Nu{\~n}ez}, Consuelo and {Phillips}, Chris},
        title = "{The Host Galaxies and Progenitors of Fast Radio Bursts Localized with the Australian Square Kilometre Array Pathfinder}",
      journal = {\apjl},
         year = 2020,
        month = jun,
       volume = {895},
       number = {2},
          eid = {L37},
        pages = {L37},
          doi = {10.3847/2041-8213/ab672e},
archivePrefix = {arXiv},
       eprint = {2005.13160},
 primaryClass = {astro-ph.GA},
       adsurl = {https://ui.adsabs.harvard.edu/abs/2020ApJ...895L..37B}
}

@ARTICLE{Petroff22,
       author = {{Petroff}, E. and {Hessels}, J.~W.~T. and {Lorimer}, D.~R.},
        title = "{Fast radio bursts at the dawn of the 2020s}",
      journal = {\aapr},
         year = 2022,
        month = dec,
       volume = {30},
       number = {1},
          eid = {2},
        pages = {2},
          doi = {10.1007/s00159-022-00139-w},
archivePrefix = {arXiv},
       eprint = {2107.10113},
 primaryClass = {astro-ph.HE},
       adsurl = {https://ui.adsabs.harvard.edu/abs/2022A&ARv..30....2P}
}

@ARTICLE{Metzger17,
       author = {{Metzger}, Brian D. and {Berger}, Edo and {Margalit}, Ben},
        title = "{Millisecond Magnetar Birth Connects FRB 121102 to Superluminous Supernovae and Long-duration Gamma-Ray Bursts}",
      journal = {\apj},
         year = 2017,
        month = may,
       volume = {841},
       number = {1},
          eid = {14},
        pages = {14},
          doi = {10.3847/1538-4357/aa633d},
archivePrefix = {arXiv},
       eprint = {1701.02370},
 primaryClass = {astro-ph.HE},
       adsurl = {https://ui.adsabs.harvard.edu/abs/2017ApJ...841...14M}
}

@ARTICLE{Margalit19,
       author = {{Margalit}, Ben and {Berger}, Edo and {Metzger}, Brian D.},
        title = "{Fast Radio Bursts from Magnetars Born in Binary Neutron Star Mergers and Accretion Induced Collapse}",
      journal = {\apj},
         year = 2019,
        month = dec,
       volume = {886},
       number = {2},
          eid = {110},
        pages = {110},
          doi = {10.3847/1538-4357/ab4c31},
archivePrefix = {arXiv},
       eprint = {1907.00016},
 primaryClass = {astro-ph.HE},
       adsurl = {https://ui.adsabs.harvard.edu/abs/2019ApJ...886..110M}
}

@ARTICLE{Kashiyama13,
       author = {{Kashiyama}, Kazumi and {Ioka}, Kunihito and {M{\'e}sz{\'a}ros}, Peter},
        title = "{Cosmological Fast Radio Bursts from Binary White Dwarf Mergers}",
      journal = {\apjl},
         year = 2013,
        month = oct,
       volume = {776},
       number = {2},
          eid = {L39},
        pages = {L39},
          doi = {10.1088/2041-8205/776/2/L39},
archivePrefix = {arXiv},
       eprint = {1307.7708},
 primaryClass = {astro-ph.HE},
       adsurl = {https://ui.adsabs.harvard.edu/abs/2013ApJ...776L..39K}
}

@ARTICLE{Tendulkar17,
       author = {{Tendulkar}, S.~P. and {Bassa}, C.~G. and {Cordes}, J.~M. and {Bower}, G.~C. and {Law}, C.~J. and {Chatterjee}, S. and {Adams}, E.~A.~K. and {Bogdanov}, S. and {Burke-Spolaor}, S. and {Butler}, B.~J. and {Demorest}, P. and {Hessels}, J.~W.~T. and {Kaspi}, V.~M. and {Lazio}, T.~J.~W. and {Maddox}, N. and {Marcote}, B. and {McLaughlin}, M.~A. and {Paragi}, Z. and {Ransom}, S.~M. and {Scholz}, P. and {Seymour}, A. and {Spitler}, L.~G. and {van Langevelde}, H.~J. and {Wharton}, R.~S.},
        title = "{The Host Galaxy and Redshift of the Repeating Fast Radio Burst FRB 121102}",
      journal = {\apjl},
         year = 2017,
        month = jan,
       volume = {834},
       number = {2},
          eid = {L7},
        pages = {L7},
          doi = {10.3847/2041-8213/834/2/L7},
archivePrefix = {arXiv},
       eprint = {1701.01100},
 primaryClass = {astro-ph.HE},
       adsurl = {https://ui.adsabs.harvard.edu/abs/2017ApJ...834L...7T}
}

@ARTICLE{Bhandari22,
       author = {{Bhandari}, Shivani and {Heintz}, Kasper E. and {Aggarwal}, Kshitij and {Marnoch}, Lachlan and {Day}, Cherie K. and {Sydnor}, Jessica and {Burke-Spolaor}, Sarah and {Law}, Casey J. and {Xavier Prochaska}, J. and {Tejos}, Nicolas and {Bannister}, Keith W. and {Butler}, Bryan J. and {Deller}, Adam T. and {Ekers}, R.~D. and {Flynn}, Chris and {Fong}, Wen-fai and {James}, Clancy W. and {Lazio}, T. Joseph W. and {Luo}, Rui and {Mahony}, Elizabeth K. and {Ryder}, Stuart D. and {Sadler}, Elaine M. and {Shannon}, Ryan M. and {Han}, JinLin and {Lee}, Kejia and {Zhang}, Bing},
        title = "{Characterizing the Fast Radio Burst Host Galaxy Population and its Connection to Transients in the Local and Extragalactic Universe}",
      journal = {\aj},
         year = 2022,
        month = feb,
       volume = {163},
       number = {2},
          eid = {69},
        pages = {69},
          doi = {10.3847/1538-3881/ac3aec},
archivePrefix = {arXiv},
       eprint = {2108.01282},
 primaryClass = {astro-ph.HE},
       adsurl = {https://ui.adsabs.harvard.edu/abs/2022AJ....163...69B}
}

@ARTICLE{Usov92,
       author = {{Usov}, V.~V.},
        title = "{Millisecond pulsars with extremely strong magnetic fields as a cosmological source of {\ensuremath{\gamma}}-ray bursts}",
      journal = {\nat},
         year = 1992,
        month = jun,
       volume = {357},
       number = {6378},
        pages = {472-474},
          doi = {10.1038/357472a0},
       adsurl = {https://ui.adsabs.harvard.edu/abs/1992Natur.357..472U}
}

@ARTICLE{Shen12,
       author = {{Shen}, Ken J. and {Bildsten}, Lars and {Kasen}, Daniel and {Quataert}, Eliot},
        title = "{The Long-term Evolution of Double White Dwarf Mergers}",
      journal = {\apj},
         year = 2012,
        month = mar,
       volume = {748},
       number = {1},
          eid = {35},
        pages = {35},
          doi = {10.1088/0004-637X/748/1/35},
archivePrefix = {arXiv},
       eprint = {1108.4036},
 primaryClass = {astro-ph.HE},
       adsurl = {https://ui.adsabs.harvard.edu/abs/2012ApJ...748...35S}
}

@ARTICLE{Ravi19,
       author = {{Ravi}, Vikram},
        title = "{The prevalence of repeating fast radio bursts}",
      journal = {Nature Astronomy},
         year = 2019,
        month = jul,
       volume = {3},
        pages = {928-931},
          doi = {10.1038/s41550-019-0831-y},
archivePrefix = {arXiv},
       eprint = {1907.06619},
 primaryClass = {astro-ph.HE},
       adsurl = {https://ui.adsabs.harvard.edu/abs/2019NatAs...3..928R}
}

@ARTICLE{Bhandari18,
       author = {{Bhandari}, S. and {Keane}, E.~F. and {Barr}, E.~D. and {Jameson}, A. and {Petroff}, E. and {Johnston}, S. and {Bailes}, M. and {Bhat}, N.~D.~R. and {Burgay}, M. and {Burke-Spolaor}, S. and {Caleb}, M. and {Eatough}, R.~P. and {Flynn}, C. and {Green}, J.~A. and {Jankowski}, F. and {Kramer}, M. and {Krishnan}, V. Venkatraman and {Morello}, V. and {Possenti}, A. and {Stappers}, B. and {Tiburzi}, C. and {van Straten}, W. and {Andreoni}, I. and {Butterley}, T. and {Chandra}, P. and {Cooke}, J. and {Corongiu}, A. and {Coward}, D.~M. and {Dhillon}, V.~S. and {Dodson}, R. and {Hardy}, L.~K. and {Howell}, E.~J. and {Jaroenjittichai}, P. and {Klotz}, A. and {Littlefair}, S.~P. and {Marsh}, T.~R. and {Mickaliger}, M. and {Muxlow}, T. and {Perrodin}, D. and {Pritchard}, T. and {Sawangwit}, U. and {Terai}, T. and {Tominaga}, N. and {Torne}, P. and {Totani}, T. and {Trois}, A. and {Turpin}, D. and {Niino}, Y. and {Wilson}, R.~W. and {Albert}, A. and {Andr{\'e}}, M. and {Anghinolfi}, M. and {Anton}, G. and {Ardid}, M. and {Aubert}, J. -J. and {Avgitas}, T. and {Baret}, B. and {Barrios-Mart{\'\i}}, J. and {Basa}, S. and {Belhorma}, B. and {Bertin}, V. and {Biagi}, S. and {Bormuth}, R. and {Bourret}, S. and {Bouwhuis}, M.~C. and {Br{\^a}nza{\c{s}}}, H. and {Bruijn}, R. and {Brunner}, J. and {Busto}, J. and {Capone}, A. and {Caramete}, L. and {Carr}, J. and {Celli}, S. and {Moursli}, R. Cherkaoui El and {Chiarusi}, T. and {Circella}, M. and {Coelho}, J.~A.~B. and {Coleiro}, A. and {Coniglione}, R. and {Costantini}, H. and {Coyle}, P. and {Creusot}, A. and {D{\'\i}az}, A.~F. and {Deschamps}, A. and {De Bonis}, G. and {Distefano}, C. and {Palma}, I. Di and {Domi}, A. and {Donzaud}, C. and {Dornic}, D. and {Drouhin}, D. and {Eberl}, T. and {Bojaddaini}, I. El and {Khayati}, N. El and {Els{\"a}sser}, D. and {Enzenh{\"o}fer}, A. and {Ettahiri}, A. and {Fassi}, F. and {Felis}, I. and {Fusco}, L.~A. and {Gay}, P. and {Giordano}, V. and {Glotin}, H. and {Gregoire}, T. and {Gracia-Ruiz}, R. and {Graf}, K. and {Hallmann}, S. and {van Haren}, H. and {Heijboer}, A.~J. and {Hello}, Y. and {Hern{\'a}ndez-Rey}, J.~J. and {H{\"o}{\ss}l}, J. and {Hofest{\"a}dt}, J. and {Hugon}, C. and {Illuminati}, G. and {James}, C.~W. and {de Jong}, M. and {Jongen}, M. and {Kadler}, M. and {Kalekin}, O. and {Katz}, U. and {Kie{\ss}ling}, D. and {Kouchner}, A. and {Kreter}, M. and {Kreykenbohm}, I. and {Kulikovskiy}, V. and {Lachaud}, C. and {Lahmann}, R. and {Lef{\`e}vre}, D. and {Leonora}, E. and {Loucatos}, S. and {Marcelin}, M. and {Margiotta}, A. and {Marinelli}, A. and {Mart{\'\i}nez-Mora}, J.~A. and {Mele}, R. and {Melis}, K. and {Michael}, T. and {Migliozzi}, P. and {Moussa}, A. and {Navas}, S. and {Nezri}, E. and {Organokov}, M. and {P{\v{a}}v{\v{a}}la{\c{s}}}, G.~E. and {Pellegrino}, C. and {Perrina}, C. and {Piattelli}, P. and {Popa}, V. and {Pradier}, T. and {Quinn}, L. and {Racca}, C. and {Riccobene}, G. and {S{\'a}nchez-Losa}, A. and {Salda{\~n}a}, M. and {Salvadori}, I. and {Samtleben}, D.~F.~E. and {Sanguineti}, M. and {Sapienza}, P. and {Sch{\"u}ssler}, F. and {Sieger}, C. and {Spurio}, M. and {Stolarczyk}, Th and {Taiuti}, M. and {Tayalati}, Y. and {Trovato}, A. and {Turpin}, D. and {T{\"o}nnis}, C. and {Vallage}, B. and {Van Elewyck}, V. and {Versari}, F. and {Vivolo}, D. and {Vizzocca}, A. and {Wilms}, J. and {Zornoza}, J.~D. and {Z{\'u}{\~n}iga}, J.},
        title = "{The SUrvey for Pulsars and Extragalactic Radio Bursts - II. New FRB discoveries and their follow-up}",
      journal = {\mnras},
         year = 2018,
        month = apr,
       volume = {475},
       number = {2},
        pages = {1427-1446},
          doi = {10.1093/mnras/stx3074},
archivePrefix = {arXiv},
       eprint = {1711.08110},
 primaryClass = {astro-ph.HE},
       adsurl = {https://ui.adsabs.harvard.edu/abs/2018MNRAS.475.1427B}
}

@ARTICLE{Meng25,
       author = {{Meng}, Min and {Deng}, Can-Min},
        title = "{Constraints on the progenitor models of fast radio bursts from population synthesis with the first CHIME/FRB catalog}",
      journal = {\aap},
         year = 2025,
        month = jun,
       volume = {698},
          eid = {A127},
        pages = {A127},
          doi = {10.1051/0004-6361/202451250},
archivePrefix = {arXiv},
       eprint = {2506.04986},
 primaryClass = {astro-ph.HE},
       adsurl = {https://ui.adsabs.harvard.edu/abs/2025A&A...698A.127M}
}

@ARTICLE{Madau17,
       author = {{Madau}, Piero and {Fragos}, Tassos},
        title = "{Radiation Backgrounds at Cosmic Dawn: X-Rays from Compact Binaries}",
      journal = {\apj},
         year = 2017,
        month = may,
       volume = {840},
       number = {1},
          eid = {39},
        pages = {39},
          doi = {10.3847/1538-4357/aa6af9},
archivePrefix = {arXiv},
       eprint = {1606.07887},
 primaryClass = {astro-ph.GA},
       adsurl = {https://ui.adsabs.harvard.edu/abs/2017ApJ...840...39M}
}

@ARTICLE{Zhang24,
       author = {{Zhang}, Ji-Guo and {Li}, Yichao and {Zou}, Jia-Ming and {Zhao}, Ze-Wei and {Zhang}, Jing-Fei and {Zhang}, Xin},
        title = "{Fast Radio Burst Energy Function in the Presence of DM$_{host}$ Variation}",
      journal = {Universe},
         year = 2024,
        month = may,
       volume = {10},
       number = {5},
          eid = {207},
        pages = {207},
          doi = {10.3390/universe10050207},
archivePrefix = {arXiv},
       eprint = {2303.16775},
 primaryClass = {astro-ph.HE},
       adsurl = {https://ui.adsabs.harvard.edu/abs/2024Univ...10..207Z}
}

@INPROCEEDINGS{DSA-2000,
       author = {{Hallinan}, Gregg and {Ravi}, V. and {Weinreb}, S. and {Kocz}, J. and {Huang}, Y. and {Woody}, D.~P. and {Lamb}, J. and {D'Addario}, L. and {Catha}, M. and {Law}, C. and {Kulkarni}, S.~R. and {Phinney}, E.~S. and {Eastwood}, M.~W. and {Bouman}, K. and {McLaughlin}, M. and {Ransom}, S. and {Siemens}, X. and {Cordes}, J. and {Lynch}, R. and {Kaplan}, D. and {Brazier}, A. and {Bhatnagar}, S. and {Myers}, S. and {Walter}, F. and {Gaensler}, B.},
        title = "{The DSA-2000 {\textemdash} A Radio Survey Camera}",
    booktitle = {Bulletin of the American Astronomical Society},
         year = 2019,
       volume = {51},
        month = sep,
          eid = {255},
        pages = {255},
          doi = {10.48550/arXiv.1907.07648},
archivePrefix = {arXiv},
       eprint = {1907.07648},
 primaryClass = {astro-ph.IM},
       adsurl = {https://ui.adsabs.harvard.edu/abs/2019BAAS...51g.255H}
}

@ARTICLE{Tang23,
       author = {{Tang}, Li and {Lin}, Hai-Nan and {Li}, Xin},
        title = "{Inferring redshift and energy distributions of fast radio bursts from the first CHIME/FRB catalog}",
      journal = {Chinese Physics C},
         year = 2023,
        month = aug,
       volume = {47},
       number = {8},
          eid = {085105},
        pages = {085105},
          doi = {10.1088/1674-1137/acda1c},
archivePrefix = {arXiv},
       eprint = {2305.19692},
 primaryClass = {astro-ph.HE},
       adsurl = {https://ui.adsabs.harvard.edu/abs/2023ChPhC..47h5105T}
}

@ARTICLE{Lei25,
       author = {{Lei}, Qing-Zhen and {Wang}, Xin-Zhe and {Deng}, Can-Min},
        title = "{A Comprehensive Study of the Energy and Redshift Distributions of the Fast Radio Burst Population Based on the First CHIME/FRB Catalog}",
      journal = {arXiv e-prints},
         year = 2025,
        month = jul,
          eid = {arXiv:2507.23122},
        pages = {arXiv:2507.23122},
          doi = {10.48550/arXiv.2507.23122},
archivePrefix = {arXiv},
       eprint = {2507.23122},
 primaryClass = {astro-ph.HE},
       adsurl = {https://ui.adsabs.harvard.edu/abs/2025arXiv250723122L}
}

@ARTICLE{Hashimoto22,
       author = {{Hashimoto}, Tetsuya and {Goto}, Tomotsugu and {Chen}, Bo Han and {Ho}, Simon C. -C. and {Hsiao}, Tiger Y. -Y. and {Wong}, Yi Hang Valerie and {On}, Alvina Y.~L. and {Kim}, Seong Jin and {Kilerci-Eser}, Ece and {Huang}, Kai-Chun and {Santos}, Daryl Joe D. and {Yamasaki}, Shotaro},
        title = "{Energy functions of fast radio bursts derived from the first CHIME/FRB catalogue}",
      journal = {\mnras},
         year = 2022,
        month = apr,
       volume = {511},
       number = {2},
        pages = {1961-1976},
          doi = {10.1093/mnras/stac065},
archivePrefix = {arXiv},
       eprint = {2201.03574},
 primaryClass = {astro-ph.HE},
       adsurl = {https://ui.adsabs.harvard.edu/abs/2022MNRAS.511.1961H}
}

@ARTICLE{Rao25,
       author = {{Rao}, Aryamann and {Ye}, Claire S. and {Fishbach}, Maya},
        title = "{Predicting the Rate of Fast Radio Bursts in Globular Clusters from Binary Black Hole Observations}",
      journal = {\apjl},
         year = 2025,
        month = jan,
       volume = {979},
       number = {1},
          eid = {L12},
        pages = {L12},
          doi = {10.3847/2041-8213/ad9f2e},
archivePrefix = {arXiv},
       eprint = {2409.20564},
 primaryClass = {astro-ph.HE},
       adsurl = {https://ui.adsabs.harvard.edu/abs/2025ApJ...979L..12R}
}

@ARTICLE{Sharma24,
       author = {{Sharma}, Kritti and {Ravi}, Vikram and {Connor}, Liam and {Law}, Casey and {Ocker}, Stella Koch and {Sherman}, Myles and {Kosogorov}, Nikita and {Faber}, Jakob and {Hallinan}, Gregg and {Harnach}, Charlie and {Hellbourg}, Greg and {Hobbs}, Rick and {Hodge}, David and {Hodges}, Mark and {Lamb}, James and {Rasmussen}, Paul and {Somalwar}, Jean and {Weinreb}, Sander and {Woody}, David and {Leja}, Joel and {Anand}, Shreya and {Das}, Kaustav Kashyap and {Qin}, Yu-Jing and {Rose}, Sam and {Dong}, Dillon Z. and {Miller}, Jessie and {Yao}, Yuhan},
        title = "{Preferential occurrence of fast radio bursts in massive star-forming galaxies}",
      journal = {\nat},
         year = 2024,
        month = nov,
       volume = {635},
       number = {8037},
        pages = {61-66},
          doi = {10.1038/s41586-024-08074-9},
archivePrefix = {arXiv},
       eprint = {2409.16964},
 primaryClass = {astro-ph.HE},
       adsurl = {https://ui.adsabs.harvard.edu/abs/2024Natur.635...61S}
}

@ARTICLE{Chatterjee17,
       author = {{Chatterjee}, S. and {Law}, C.~J. and {Wharton}, R.~S. and {Burke-Spolaor}, S. and {Hessels}, J.~W.~T. and {Bower}, G.~C. and {Cordes}, J.~M. and {Tendulkar}, S.~P. and {Bassa}, C.~G. and {Demorest}, P. and {Butler}, B.~J. and {Seymour}, A. and {Scholz}, P. and {Abruzzo}, M.~W. and {Bogdanov}, S. and {Kaspi}, V.~M. and {Keimpema}, A. and {Lazio}, T.~J.~W. and {Marcote}, B. and {McLaughlin}, M.~A. and {Paragi}, Z. and {Ransom}, S.~M. and {Rupen}, M. and {Spitler}, L.~G. and {van Langevelde}, H.~J.},
        title = "{A direct localization of a fast radio burst and its host}",
      journal = {\nat},
         year = 2017,
        month = jan,
       volume = {541},
       number = {7635},
        pages = {58-61},
          doi = {10.1038/nature20797},
archivePrefix = {arXiv},
       eprint = {1701.01098},
 primaryClass = {astro-ph.HE},
       adsurl = {https://ui.adsabs.harvard.edu/abs/2017Natur.541...58C}
}

@ARTICLE{Marcote17,
       author = {{Marcote}, B. and {Paragi}, Z. and {Hessels}, J.~W.~T. and {Keimpema}, A. and {van Langevelde}, H.~J. and {Huang}, Y. and {Bassa}, C.~G. and {Bogdanov}, S. and {Bower}, G.~C. and {Burke-Spolaor}, S. and {Butler}, B.~J. and {Campbell}, R.~M. and {Chatterjee}, S. and {Cordes}, J.~M. and {Demorest}, P. and {Garrett}, M.~A. and {Ghosh}, T. and {Kaspi}, V.~M. and {Law}, C.~J. and {Lazio}, T.~J.~W. and {McLaughlin}, M.~A. and {Ransom}, S.~M. and {Salter}, C.~J. and {Scholz}, P. and {Seymour}, A. and {Siemion}, A. and {Spitler}, L.~G. and {Tendulkar}, S.~P. and {Wharton}, R.~S.},
        title = "{The Repeating Fast Radio Burst FRB 121102 as Seen on Milliarcsecond Angular Scales}",
      journal = {\apjl},
         year = 2017,
        month = jan,
       volume = {834},
       number = {2},
          eid = {L8},
        pages = {L8},
          doi = {10.3847/2041-8213/834/2/L8},
archivePrefix = {arXiv},
       eprint = {1701.01099},
 primaryClass = {astro-ph.HE},
       adsurl = {https://ui.adsabs.harvard.edu/abs/2017ApJ...834L...8M}
}

@ARTICLE{Marcote20,
       author = {{Marcote}, B. and {Nimmo}, K. and {Hessels}, J.~W.~T. and {Tendulkar}, S.~P. and {Bassa}, C.~G. and {Paragi}, Z. and {Keimpema}, A. and {Bhardwaj}, M. and {Karuppusamy}, R. and {Kaspi}, V.~M. and {Law}, C.~J. and {Michilli}, D. and {Aggarwal}, K. and {Andersen}, B. and {Archibald}, A.~M. and {Bandura}, K. and {Bower}, G.~C. and {Boyle}, P.~J. and {Brar}, C. and {Burke-Spolaor}, S. and {Butler}, B.~J. and {Cassanelli}, T. and {Chawla}, P. and {Demorest}, P. and {Dobbs}, M. and {Fonseca}, E. and {Giri}, U. and {Good}, D.~C. and {Gourdji}, K. and {Josephy}, A. and {Kirichenko}, A. Yu. and {Kirsten}, F. and {Landecker}, T.~L. and {Lang}, D. and {Lazio}, T.~J.~W. and {Li}, D.~Z. and {Lin}, H. -H. and {Linford}, J.~D. and {Masui}, K. and {Mena-Parra}, J. and {Naidu}, A. and {Ng}, C. and {Patel}, C. and {Pen}, U. -L. and {Pleunis}, Z. and {Rafiei-Ravandi}, M. and {Rahman}, M. and {Renard}, A. and {Scholz}, P. and {Siegel}, S.~R. and {Smith}, K.~M. and {Stairs}, I.~H. and {Vanderlinde}, K. and {Zwaniga}, A.~V.},
        title = "{A repeating fast radio burst source localized to a nearby spiral galaxy}",
      journal = {\nat},
         year = 2020,
        month = jan,
       volume = {577},
       number = {7789},
        pages = {190-194},
          doi = {10.1038/s41586-019-1866-z},
archivePrefix = {arXiv},
       eprint = {2001.02222},
 primaryClass = {astro-ph.HE},
       adsurl = {https://ui.adsabs.harvard.edu/abs/2020Natur.577..190M}
}

@ARTICLE{Fong21,
       author = {{Fong}, Wen-fai and {Dong}, Yuxin and {Leja}, Joel and {Bhandari}, Shivani and {Day}, Cherie K. and {Deller}, Adam T. and {Kumar}, Pravir and {Prochaska}, J. Xavier and {Scott}, Danica R. and {Bannister}, Keith W. and {Eftekhari}, Tarraneh and {Gordon}, Alexa C. and {Heintz}, Kasper E. and {James}, Clancy W. and {Kilpatrick}, Charles D. and {Mahony}, Elizabeth K. and {Rouco Escorial}, Alicia and {Ryder}, Stuart D. and {Shannon}, Ryan M. and {Tejos}, Nicolas},
        title = "{Chronicling the Host Galaxy Properties of the Remarkable Repeating FRB 20201124A}",
      journal = {\apjl},
         year = 2021,
        month = oct,
       volume = {919},
       number = {2},
          eid = {L23},
        pages = {L23},
          doi = {10.3847/2041-8213/ac242b},
archivePrefix = {arXiv},
       eprint = {2106.11993},
 primaryClass = {astro-ph.GA},
       adsurl = {https://ui.adsabs.harvard.edu/abs/2021ApJ...919L..23F}
}

@ARTICLE{Niu22,
       author = {{Niu}, C. -H. and {Aggarwal}, K. and {Li}, D. and {Zhang}, X. and {Chatterjee}, S. and {Tsai}, C. -W. and {Yu}, W. and {Law}, C.~J. and {Burke-Spolaor}, S. and {Cordes}, J.~M. and {Zhang}, Y. -K. and {Ocker}, S.~K. and {Yao}, J. -M. and {Wang}, P. and {Feng}, Y. and {Niino}, Y. and {Bochenek}, C. and {Cruces}, M. and {Connor}, L. and {Jiang}, J. -A. and {Dai}, S. and {Luo}, R. and {Li}, G. -D. and {Miao}, C. -C. and {Niu}, J. -R. and {Anna-Thomas}, R. and {Sydnor}, J. and {Stern}, D. and {Wang}, W. -Y. and {Yuan}, M. and {Yue}, Y. -L. and {Zhou}, D. -J. and {Yan}, Z. and {Zhu}, W. -W. and {Zhang}, B.},
        title = "{A repeating fast radio burst associated with a persistent radio source}",
      journal = {\nat},
         year = 2022,
        month = jun,
       volume = {606},
       number = {7916},
        pages = {873-877},
          doi = {10.1038/s41586-022-04755-5},
archivePrefix = {arXiv},
       eprint = {2110.07418},
 primaryClass = {astro-ph.HE},
       adsurl = {https://ui.adsabs.harvard.edu/abs/2022Natur.606..873N}
}

\end{document}